\documentclass[journal]{IEEEtran}

\ifCLASSINFOpdf
  \usepackage[pdftex]{graphicx}
  \graphicspath{{./Figures/}{./photos/}}
  \DeclareGraphicsExtensions{.pdf,.png,.jpg}
\else
  \usepackage[dvips]{graphicx}
  \graphicspath{{./Figures/}{./photos/}}
  \DeclareGraphicsExtensions{.pdf,.png}
\fi

\usepackage{amsmath}
\ifCLASSOPTIONcompsoc
 \usepackage[caption=false,font=normalsize,labelfont=sf,textfont=sf]{subfig}
\else
 \usepackage[caption=false,font=footnotesize]{subfig}
\fi
\usepackage{dblfloatfix}
\usepackage{url}

\usepackage{amsfonts}

\usepackage{flushend}

\usepackage{enumitem}

\usepackage{multirow}

\newcommand{\sgn}{\textrm{sgn}}
\newcommand\given[1][]{\:#1\vert\:}

\DeclareMathOperator*{\arginf}{arg\,inf}

\usepackage{algorithm}
\usepackage{algorithmic}
\usepackage{siunitx}

\makeatletter
\newcommand\fs@norules{\def\@fs@cfont{\bfseries}\let\@fs@capt\floatc@ruled
  \def\@fs@pre{}%
  \def\@fs@post{}%
  \def\@fs@mid{\kern3pt}%
  \let\@fs@iftopcapt\iftrue}
\makeatother
\floatstyle{norules}
\restylefloat{algorithm}

\begin{document}

%
\title{Constructing High-Order Signed Distance Maps from Computed Tomography Data with Application to Bone Morphometry}
%
%
%


\author{
  Bryce~A.~Besler, 
  Tannis~D.~Kemp,
  Nils~D.~Forkert, 
  Steven~K.~Boyd

\thanks{This work was supported by the Natural Sciences and Engineering Research Council (NSERC) of Canada, grant RGPIN-2019-4135.}
\thanks{B.A. Besler and T. D. Kemp are in the McCaig Institute for Bone and Joint Health, University of Calgary Canada. N.D. Forkert is with the Department of Radiology and the Hotchkiss Brain Institute, University of Calgary, Canada. S.K. Boyd is with the Department of Radiology and McCaig Institute for Bone and Joint Health, University of Calgary Canada e-mail: skboyd@ucalgary.ca}

}

%
%

\markboth{}%
{Besler \MakeLowercase{\textit{et al.}}: Constructing High-Order Signed Distance Maps from Computed Tomography}
%



\maketitle

\begin{abstract}
  An algorithm is presented for constructing high-order signed distance fields for two phase materials imaged with computed tomography.
  The signed distance field is high-order in that it is free of the quantization artifact associated with the distance transform of sampled signals.
  The narrowband is solved using a closest point algorithm extended for implicit embeddings that are not a signed distance field.
  The high-order fast sweeping algorithm is used to extend the narrowband to the remainder of the domain.
  The order of accuracy of the narrowband and extension methods are verified on ideal implicit surfaces.
  The method is applied to ten excised cubes of bovine trabecular bone.
  Localization of the surface, estimation of phase densities, and local morphometry is validated with these subjects.
  Since the embedding is high-order, gradients and thus curvatures can be accurately estimated locally in the image data.
\end{abstract}


%
\IEEEpeerreviewmaketitle

\section{Introduction}
\label{sec:introduction}
This paper develops a method of constructing signed distance fields from computed tomography datasets.
Primarily, the method is concerned with two-phase materials such as trabecular bone, synthetic biomaterials, and soil samples.
Being biphasic defines the interface between the two materials as an orientable surface, advantageous for modelling changes over time and measuring metrics of the surface such as volume, area, curvatures, and thickness.

Classically, a signed distance field would be constructed in two steps.
First, an appropriate segmentation method would generate a binary image defining the object in space.
Second, the exact signed distance transform would produce the distance map from the binary image~\cite{breu1995linear}.
However, the exact signed distance transform computes a quantized version of the true signed distance field~\cite{besler2020artifacts}.
This produces errors for geometric flow and curvature computation algorithms because the algorithm is only initialized first-order accurate, resulting in large errors in measurements and curvature computation.
Furthermore, constructing high-order signed distance maps from binary sampled signals is a non-unique problem~\cite{besler2021highorder}, permitting multiple embeddings that satisfy the same binary image.

The task presented here is to construct the signed distance field with a high-order of accuracy from data measured using computed tomography.
The field will be constructed directly from grayscale data where gradients are well-defined.
Once constructed, the densities of the respective phases can be estimated.
Finally, it is demonstrated that morphometry can be performed accurately on these embeddings, a task that is too noisy when computed from the exact signed distance transform.

\section{Method}
\label{sec:method}

\subsection{Problem Definition}
\label{subsec:problem_definition}
The problem is to construct a signed distance field, $\phi$, from a density-calibrated computed tomography image, $\rho$.
Both will be represented on a sampled, rectilinear n-dimensional grid $\Omega \subset \mathbf{Z}^n$.
Let the domain $\Omega$ be split between the two phases, $\Omega_1$ and $\Omega_2$ such that $\Omega = \Omega_1 \cup \Omega_2$.
Associate to each phase a density, $\rho_1$ and $\rho_2$.
$\Gamma$ is the interface between the two materials.
We represent the interface implicitly as the zero level set of an embedding function $u$.
\begin{equation}
  \Gamma = \left\{x \given u(x) = 0 \right\}
\end{equation}
If the embedding is known, the density field can be reconstructed knowing the density of the two phases.
\begin{equation}
  \label{eqn:density_construction}
  \rho = \rho_1 \theta(-u) + \rho_2 \left[1 - \theta(-u)\right]
\end{equation}
where $\theta$ is the Heaviside function and the convention inside-is-negative is used.
This construction originates in the simulation of incompressible two-phase flow~\cite{sussman1994level}.

In many cases, we want the embedding function to satisfy an additional constraint, namely the Eikonal equation:
\begin{equation}
  | \nabla \phi | F = 1
\end{equation}
where  $\nabla$ is the gradient operator, $|\cdot|$ is the vector norm, and $F$ is a spatially varying speed function which yields the distance function when $F = 1$.
Here, we used the symbol $\phi$ to distinguish the signed distance field from any general embedding $u$.

\begin{figure*}
  \centering
  \begin{tabular}{ccccc}
    \subfloat[$\rho$]{
      \includegraphics[width=0.165\linewidth]{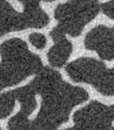}%
      \label{fig:overview:rho}
    } &
    \subfloat[$G_\sigma * \rho$]{
      \includegraphics[width=0.165\linewidth]{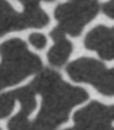}%
      \label{fig:overview:den}
    } &
    \subfloat[$\psi = T - G_\sigma * \rho$]{
      \includegraphics[width=0.165\linewidth]{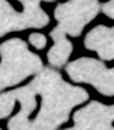}%
      \label{fig:overview:psi}
    } &
    \subfloat[$\phi_{narrowband}$]{
      \includegraphics[width=0.165\linewidth]{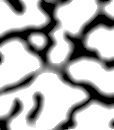}%
      \label{fig:overview:init}
    } &
    \subfloat[$\phi$]{
      \includegraphics[width=0.165\linewidth]{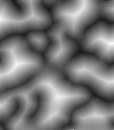}%
      \label{fig:overview:hsdt}
    }
  \end{tabular}
  \caption{Overview of computing the signed distance transform from a density image for a cube of trabecular bone. (\ref{fig:overview:rho}) the original density image is (\ref{fig:overview:den}) smoothed and (\ref{fig:overview:psi}) subtracted from a threshold to give an implicit surface. (\ref{fig:overview:init}) The narrowband is solved from the implicit surface (\ref{fig:overview:hsdt}) which is then marched outwards using the fast sweeping method.}
  \label{fig:overview}
\end{figure*}

A signed distance field is highly desirable for many reasons.
First, it can be used directly for many morphometric measures such as measuring areas, volumes, curvatures, and thicknesses~\cite{hildebrand1997new,chan2001active,besler2021morph}.
Second, the signed distance field can represent arbitrary topology surfaces and evolve them without splitting and merging rules as would be needed for parametric meshes~\cite{osher1988fronts}.
Third, during surface evolution problems, the Eikonal equation provides a way of rapidly redistancing the embedding during evolution~\cite{sussman1994level,peng1999pde}.
Constructing $\phi$ from $\rho$ is the central challenge of this paper.

\subsection{Solver}
\label{subsec:solver}
We present a method of constructing $\phi$ from $\rho$ in computed tomography datasets given the assumptions of Section~\ref{subsec:problem_definition}.
An overview of the method is given in Figure~\ref{fig:overview}.

\subsubsection{Constructing an Implicit Surface}
\label{subsubsec:implicit_surface}
An implicit representation of the surface is constructed by shifting the image by a threshold.
This causes the zero level set to be at the level of the threshold.
Standard preprocessing can also be performed on the image such as smoothing to remove noise.
A simple expression is used in this work:
\begin{equation}
  \label{eqn:implicit_surface}
  \psi = T - G_\sigma * \rho
\end{equation}
where $G_\sigma$ is a Gaussian kernel with standard deviation $\sigma$, $*$ is convolution, and $T$ is a threshold.
The symbol $\psi$ is used to denote that this is not a signed distance transform.
Importantly, this is an implicit representation of the surface $\Gamma$ that is sub-voxel.
This construction is inspired by a sub-pixel distance map method~\cite{kimmel1996sub}.

Of note, the threshold function can be made spatially varying, $T = T(x)$, important to high resolution bone scans where cortical and spongy bone compartments are often segmented with different thresholds~\cite{manske2015human}.
Similarly, alternative preprocessing filters can be used that match the specific application.

\begin{figure}[b]
  \centering
  \includegraphics[width=0.8\linewidth]{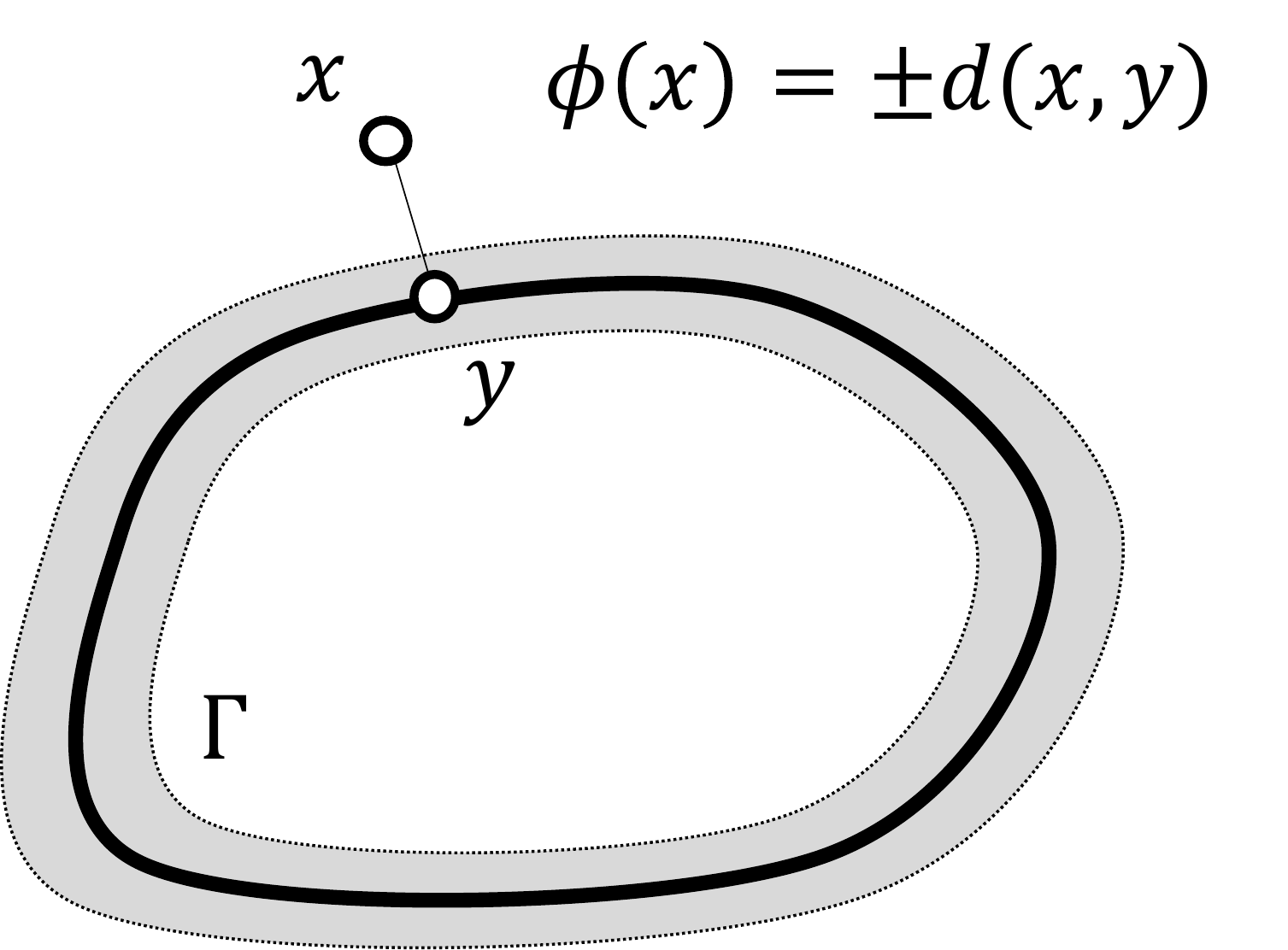}%
  \caption{Visual representation of the problem. To construct the signed distance field $\phi$, the distance to the implicit curve $\Gamma$ is found. This is done first in the narrowband (gray) and then marched outwards.}
  \label{fig:problem_definition}
\end{figure}

\subsubsection{Solving the Narrowband}
\label{subsubsec:narrowband}
A modification of the closest point method~\cite{ruuth2008simple} is used to construct the signed distance function in the narrowband.
The concept is to take each point on the rectilinear grid, $x$, that is in the narrowband and find the closest point, $y$, on the implicit surface:
\begin{equation}
  y = \arginf_{p \in \Gamma} d(p,x)
\end{equation}
where $d(\cdot,\cdot)$ is the Euclidean distance.
The distance between these two points gives the embedding function:
\begin{equation}
  \phi(x) = \pm d(x, y)
\end{equation}
where the sign is determined from the sign of $\psi(x)$.
This is demonstrated graphically in Figure~\ref{fig:problem_definition}.

The closest point method uses gradients of $\psi$ to iteratively move a point in the narrowband onto the zero level set.
However, the closest point method originally used a signed distance field, which is not available in this case.
The algorithm is modified for the embedding in Equation~\ref{eqn:implicit_surface}.

Consider a point in the narrowband.
This point is moved in the negative direction of the gradient until it is sufficiently close to the zero level set.
However, the embedding gradient points ``uphill'' (away from the medial axis) in both the positive and negative sections of the curve.
As such, the direction is modified by multiplying the sign of the embedding so both interior and exterior points move towards the zero level set.
The update equation is then:
\begin{equation}
  \label{eqn:update}
  y_{n+1} = y_n - \lambda\, \sgn(\psi(y_n))\, n(y_n)
\end{equation}
where $\lambda$ is a step size, $\sgn(\cdot)$ is the sign function, and $n(y)$ is the normal at the point $y$.
As is standard in image processing and level set methods, the normal is a function of the gradient:
\begin{equation}
  n = \frac{\nabla \psi}{|\nabla \psi|}
\end{equation}
The process is repeated till the embedding at the point $y$ is sufficiently close to zero.

\begin{figure}
  \centering
  \includegraphics[width=0.6\linewidth]{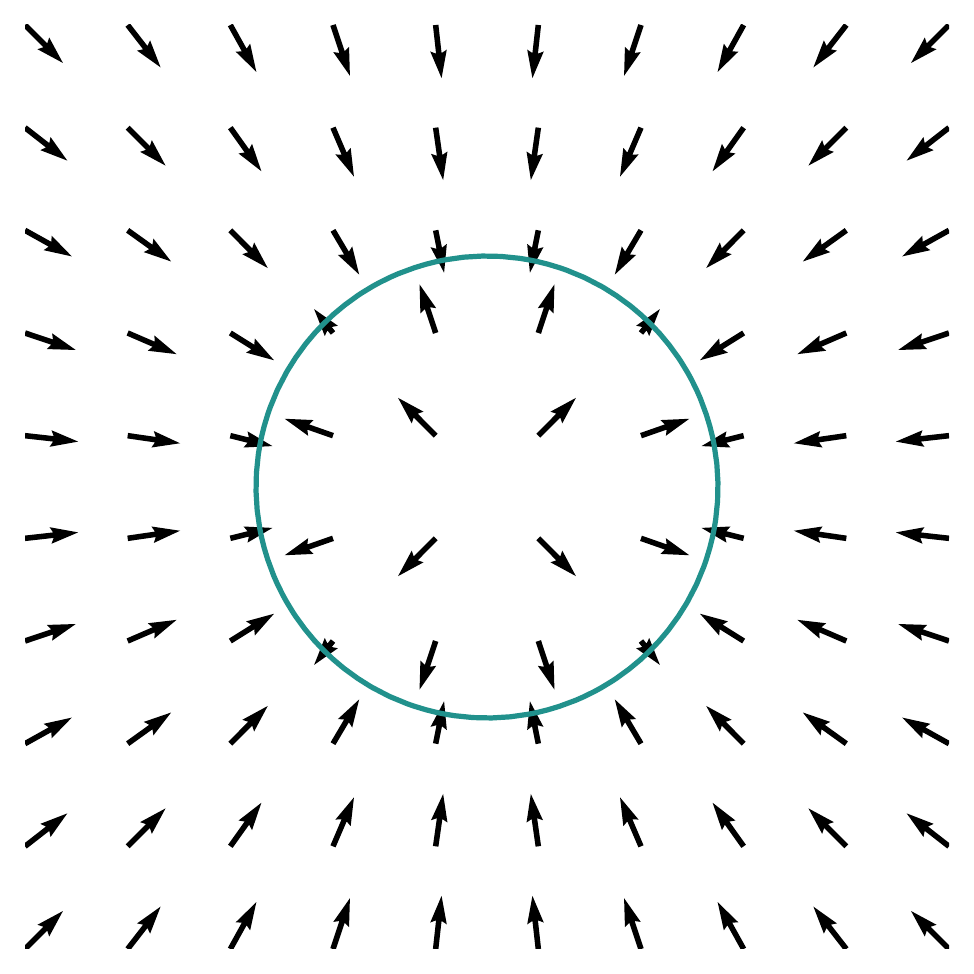}%
  \caption{Vector field, $v(x)$, for moving narrowband voxels towards the surface. The surface $\Gamma$ is shown as the gray level set. Vector length approaches zero as they get closer to the curve due to the regularized sign function.}
  \label{fig:vector}
\end{figure}

Equation~\ref{eqn:update} can be rewritten as a continuous partial differential equation:
\begin{equation}
  \label{eqn:cont_cp}
  \frac{\partial x}{\partial \tau} = \sgn(-\psi(x))\frac{\nabla \psi(x)}{|\nabla \psi(x)|}
\end{equation}
where $\tau$ is a virtual time parameter.
The closest point method is then a forward Euler approximation for the initial value integration problem where the point is integrated through a stationary vector field:
\begin{equation}
  \label{eqn:vector}
  v(x) = \sgn(-\psi(x))\frac{\nabla \psi(x)}{|\nabla \psi(x)|}
\end{equation}
This vector field is plotted in Figure~\ref{fig:vector}.

To solve Equation~\ref{eqn:cont_cp}, an interpolator is needed such that the field $\psi$ can be assessed between sample points.
We construct the interpolator using basis splines:
\begin{equation}
  \tilde{\psi}(x) = \sum_{q \in \Omega} c(q) \beta^n(x-q)
\end{equation}
where $\beta^n$ is the n\textsuperscript{th} order normalized basis spline, $c(q)$ are the coefficients of the interpolant, and $\tilde{\psi}(x)$ is the interpolating function for $\psi$.
Conveniently, once the coefficients have been constructed, derivatives can be evaluated quickly:
\begin{equation}
  \frac{\partial \tilde{\psi}(x)}{\partial x_i} = \sum_{q \in \Omega} c(q) \frac{\partial}{\partial x_i} \beta^n(x-q)
\end{equation}
This allows the computation of normals and magnitude gradients without having to compute finite differences in the image.
B-splines have been well studied~\cite{unser1993a,unser1993b,unser1999splines} and are particularly appropriate for this task.
First, they have a finite support, making them rapid to evaluate.
Second, the smoothness of the interpolating function is the order of the normalized basis spline, allowing a simple selection of the order of accuracy.
The drawback is that the construction of the basis splines can take some time and requires additional memory to store the coefficients.
However, the time cost of construction is made up by the speed of evaluation.

Combining the interpolator and vector field integrator gives the closest point algorithm presented in Algorithm~\ref{algo:cp}.
The integration time step is selected as $\lambda = h$, making the time step on the order of a voxel.
Third order basis splines are used in the interpolant for accurate computation of normals.
The integration stops when the point is sufficiently close to the implicit surface, given by $h^3$.
Since a signed distance function is not used, the stopping criterion is modified as described in Appendix~\ref{app:approx}.

A numerical approximation is made for the sign function:
\begin{equation}
  \sgn(\psi) = \frac{\psi}{\sqrt{\psi^2 + |\nabla \psi|^2 \delta^2}}
\end{equation}
where $\delta$ is a small constant preventing the denominator from being zero, selected to be the sample spacing, $h$.
The regularization constant $\delta$ is weighted by the gradient magnitude for the same reasons presented in Appendix~\ref{app:approx}.
Such a regularization has been used previously~\cite{peng1999pde}.
Regularization of the sign function has the additional benefit of causing the rate of Equation~\ref{eqn:update} to approach zero as the point $y_n$ approaches the implicit curve, preventing oscillations around the zero crossing.
For images with anisotropic spacing, $h$ is selected as the minimum of the voxel edge lengths.

\begin{algorithm}[t]
  \caption{Closest Point Algorithm, $CP_\odot$. The vector field $v$ is defined in Equation~\ref{eqn:vector} and computed using the regularized sign equation and basis spline interpolator.}
  \label{algo:cp}
  \begin{algorithmic}[1]
  \renewcommand{\algorithmicrequire}{\textbf{Input:}}
  \renewcommand{\algorithmicensure}{\textbf{Output:}}
  \REQUIRE $\psi$, $x$, $h$
  \ENSURE  $y$
  \\ \textit{Initialization}
   \STATE {$y = x$}
   \STATE {$\tilde{\psi} = \text{BuildBSpline}(\psi)$}
  \\ \textit{Process}
   \WHILE {$|\tilde{\psi}(y)| > h^3 | \nabla \tilde{\psi}(y)|$}
    \STATE {$y \leftarrow y - h\, \sgn(\tilde{\psi}(y))\, n(y)$}
   \ENDWHILE
  \RETURN $y$
  \end{algorithmic}
\end{algorithm}

An issue with Algorithm~\ref{algo:cp} is that it does not enforce collinearity with the surface~\cite{coquerelle2016fourth} where the vector from $x$ to $y$ must be collinear with the surface normal at $y$.
Even with high-order Runge-Kutta methods, integration through the vector field does not satisfy the collinearity property.
In two dimensions, the candidate point $y$ can be corrected by moving along the tangent~\cite{coquerelle2016fourth}.
However, in high dimensions, there is not a single tangent but a tangent plane.
We extend the collinearity method to arbitrary dimensions here.

\begin{figure}
  \centering
  \includegraphics[width=0.8\linewidth]{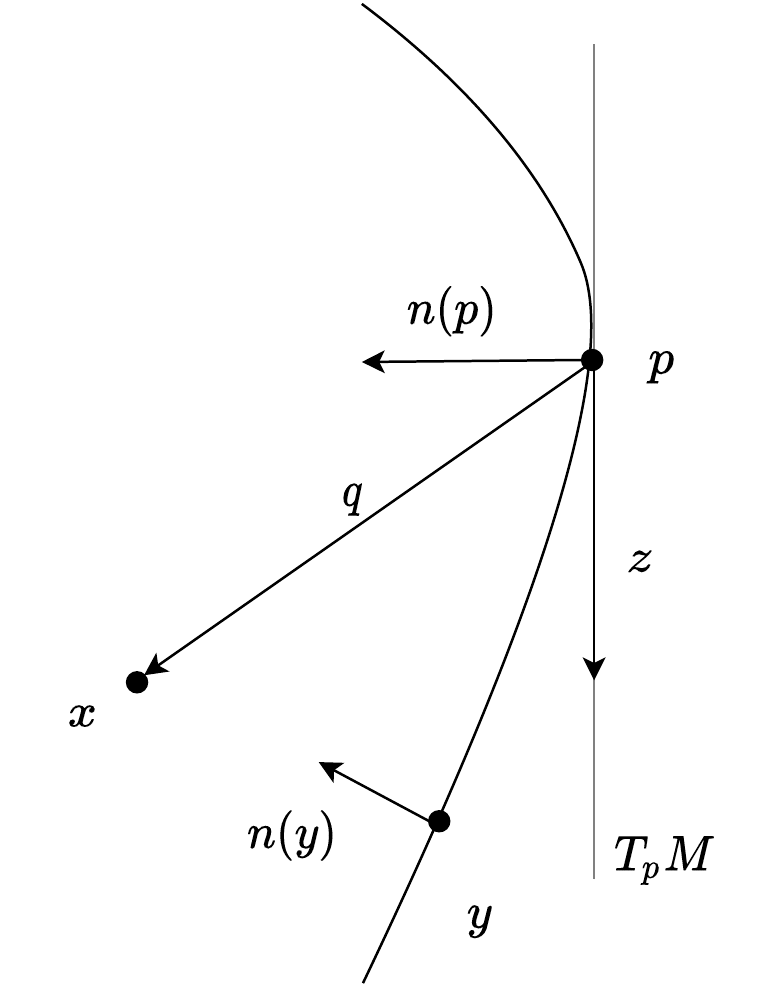}%
  \caption{Moving of a candidate point on the manifold closer to the collinear point.}
  \label{fig:projection}
\end{figure}

Let $x$ be the candidate point and define $p$ to be the current solution from $CP_\odot$.
Define the vector $q$ as pointing from $p$ to $x$.
Define $T_pM$ to be the tangent plane at the point $p$ on the manifold.
The candidate point is moved by a vector being the projection of $q$ onto $T_pM$.
\begin{eqnarray}
  z &=& \text{proj}_{T_pM}(q) \\
  z &=& q - (q \cdot n(p)) n(p)
\end{eqnarray}
This is shown graphically in Figure~\ref{fig:projection}.
The new candidate point is now the original plus this projection.
\begin{equation}
  \label{eqn:collinear_update}
  x_{n+1} = p_{n} + z
\end{equation}
The projected vector $z$ will be zero when it is collinear with the normal of the tangent plane, giving a natural stopping condition.
At areas of large curvature, Equation~\ref{eqn:collinear_update} can cause oscillations in the candidate point.
To avoid oscillations, the tangent plane vector is weighted:
\begin{equation}
  \label{eqn:collinear_update_with_stepsize}
  x_{n+1} = p_{n} + \beta z
\end{equation}
where $\beta \in [0, 1]$ is the step size, taken to be $0.5$ in this work.
The magnitude of $z$ can be understood as the size of the open ball at $p_n$ and enforcing the radius to be less than a power of the spacing enforces an order of accuracy.
This produces slower convergence in flat areas where $\beta = 1$ would be exact, but greatly improves the convergence in high curvature areas.
The surfaces of interest have very few flat locations, making this trade-off advantageous.
The order of accuracy is selected the same of $CP_\odot$, $h^3$.
This is termed the Collinear Closest Point Algorithm~\cite{coquerelle2016fourth}, $CP_\perp$, and is described in Algorithm~\ref{alg:cp_perp}.

\begin{algorithm}[b]
  \caption{Collinear Closest Point Algorithm, $CP_\perp$}
  \label{alg:cp_perp}
  \begin{algorithmic}[1]
  \renewcommand{\algorithmicrequire}{\textbf{Input:}}
  \renewcommand{\algorithmicensure}{\textbf{Output:}}
  \REQUIRE $\psi$, $x$, $h$
  \ENSURE  $y$
  \\ \textit{Initialization}
    \STATE {$y = CP_\odot(\psi, x, h)$}
    \STATE {$z = h \hat{y}$}
  \\ \textit{Process}
   \WHILE {$|z| > h^3$}
    \STATE {$z \leftarrow \text{proj}_{T_pM}(\vec{yx})$}
    \STATE {$y \leftarrow CP_\odot(\psi, y+ 0.5 z, h)$}
   \ENDWHILE
  \RETURN $y$
  \end{algorithmic}
\end{algorithm}

In rare cases, $CP_\odot$ or $CP_\perp$ will not converge.
This can occur when the initial point is at the medial axis of the image, the gradient of the edge is too flat causing noise in the normal estimation, or the noise in the density image is too large.
A maximum iteration is set (100 in this work for both algorithms) and if the maximum iterations are reached, the point is solved with the method of Section~\ref{subsubsec:extending}.
This causes a degradation of accuracy locally, but keeps the implementation stable.

\subsubsection{Extending off the Narrowband}
\label{subsubsec:extending}
After solving the narrowband, the signed distance field must be extended to the rest of the domain.
This is done using the fast sweeping method of Zhao~\cite{zhao2005fast,zhang2006high}.
The fast marching method~\cite{tsitsiklis1995efficient,sethian1996fast,chopp2001some} and fast iterative method~\cite{jeong2008fast} are also suitable algorithms for extending the narrowband, but the authors prefer the simplicity of the fast sweeping method.

As signed distance fields are non-smooth surfaces (technically, Lipschitz continuous), they cannot be better in general than first-order accurate at discontinuities~\cite{zhang2006high}, such as the medial axis where characteristics flow into each other.
However, away from the medial axis, arbitrary order of accuracy can be achieved.

The sweeping method proceeds in three steps.
First, the absolute value of the narrowband voxels are taken to remove the sign condition.
This solves for the unsigned distance field from the surface.
Next, an upwind quadratic solver is used to solve the quadratic equations at each voxel.
Importantly, as the method is upwind, the solution is propagated from the narrowband to the rest of the domain.
The solver is fast because alternating sweeping directions cascades the solution from the surface to the domain,
Finally, the sign condition is reintroduced by multiplying the solved distance field by the sign of $\psi$.
There are techniques to include the sign condition in the solver~\cite{besler2021highorder}, but they are considerably slower due to the branching requirements and are considerably more difficult to implement.

\subsection{Component Estimation}
\label{subsec:component_estimation}
Now that a non-parametric representation of the interface is available, we would like to measure the parameters $\rho_1$ and $\rho_2$.
We follow here the method of Chan and Vese~\cite{chan2001active}:
\begin{eqnarray}
  \label{eqn:rho1}
  \rho_1 &=& \frac{\int_\Omega \rho(x) \theta(-\phi(x)) dV}{\int_\Omega \theta(-\phi(x)) dV} \\
  \label{eqn:rho2}
  \rho_2 &=& \frac{\int_\Omega \rho(x) \left[1 - \theta(-\phi(x))\right] dV}{\int_\Omega \left[1 - \theta(-\phi(x))\right] dV}
\end{eqnarray}
where $\rho_1$ is the density of the phase inside the embedding and $\rho_2$ is the density of the phase outside the embedding.
In the context of trabecular bone, $\rho_1$ is the tissue mineral density and $\rho_2$ the marrow mineral density.
A nice feature of this method of estimating the phase densities is that the reconstructed density image (Equation~\ref{eqn:density_construction}) will have an average density near the original image, allowing for the computation of average density from the embedded image and the phase densities.
Of note, components are estimated from the smoothed density image, $G_\sigma * \rho$, not the implicit surface, $\psi$, or the original density image, $\rho$.
This keeps the phase density estimation and surface localization consistent with any filtering.

Numerical approximations are needed for the integration and Heaviside function.
Numerical integration is performed with Simpson's rule, particularly important for surfaces with few samples across their thickness.
A regularized version of the Heaviside function is used:
\begin{equation}
  \label{eqn:numerical_heaviside}
  \theta_\epsilon(x) = \begin{cases}
    \frac{1}{2} \left[1 + \frac{x}{\epsilon} + \frac{1}{\pi} \sin\left(\frac{\pi x}{\epsilon}\right) \right] & \lvert x \rvert \leq \epsilon\\
    1 & x > \epsilon \\
    0 & x < -\epsilon 
  \end{cases}   
\end{equation}
where $2 \epsilon$ is the finite support of the numerical Heaviside $\theta_\epsilon$.
A finite support is advantageous for structures with few samples across their thickness as it avoids extending into the medial axis.

\subsection{Narrowband Size}
\label{subsec:narrowband_size}
The voxels in the narrowband must be determined.
Since the initial embedding is not a signed distance transform, the standard absolute value threshold cannot be used:
\begin{equation}
  \Omega_{NB} = \left\{x \given \lvert \phi(x) \rvert < \gamma \right\}
\end{equation}
where $\lvert \cdot \rvert$ is the absolute value function and $\gamma$ is the narrowband size.
Instead, the set of voxels in the narrowband is determined by thresholding the embedding below zero and performing morphological operators:
\begin{eqnarray}
  I &=& \psi < 0 \\
  J &=& I \oplus S - I \ominus S
\end{eqnarray}
where $\oplus$ is morphological dilation, $\ominus$ is morphological erosion, and $S$ is a suitable structuring element.
This is considerably slower than the thresholding approach but selects a technically correct narrowband.
The cross structuring element is used in this work and the radius is set to half the stencil size minus one.
\begin{equation}
  r = \frac{\text{Stencil Size} - 1}{2}
\end{equation}
This makes the narrowband large enough that a finite difference stencil placed outside the narrowband does not extend across the narrowband.
The image $J$ then has the value one for all voxels in the narrowband and zero for all elements outside the narrowband.

\begin{figure}
  \centering
  \begin{tabular}{ccc}
    \subfloat[Sphere]{
      \includegraphics[width=0.26\linewidth]{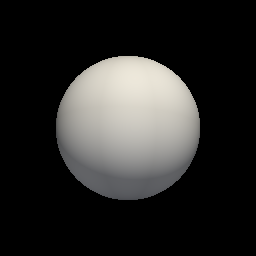}%
      \label{fig:surfaces:sphere}
    } &
    \subfloat[Torus]{
      \includegraphics[width=0.26\linewidth]{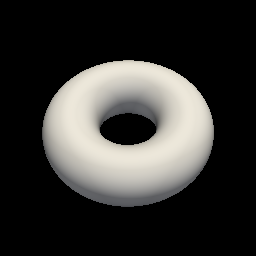}%
      \label{fig:surfaces:torus}
    } &
    \subfloat[Double Spheres]{
      \includegraphics[width=0.26\linewidth]{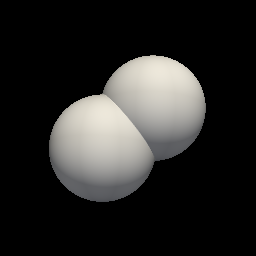}%
      \label{fig:surfaces:double_sphere}
    }
  \end{tabular}
  \caption{Visualization of the ideal surfaces for testing order of accuracy.}
  \label{fig:surfaces}
\end{figure}

\section{Experiments}
\label{sec:experiments}

\subsection{Correctness on Ideal Surfaces}
A series of experiments are performed on ideal surfaces to validate the correctness of the algorithms.
The signed distance field of a sphere and torus are known analytically:
\begin{eqnarray}
  \phi_{sphere} &=& \sqrt{x^2 + y^2 + z^2} - r \\
  \phi_{torus} &=& \sqrt{\left(\sqrt{x^2 + y^2} - c\right)^2 + z^2} - a
\end{eqnarray}
where $r$ is the radius of the sphere, $a$ is the radius of the torus, and $c$ is the distance to the center of the torus.
This allows us to construct a ground truth $\phi$, create a density image $\rho$, and assess the ability of the algorithm to recover $\phi$ from $\rho$.
Additionally, two intersecting spheres are constructed, producing a continuous but non-smooth implicit surface.
The surface parameters are $r=5$ for the sphere, $a=2$, $c=3$ for the torus, and $r=5$ for the double spheres with separation distance $3\sqrt{3}$.
Images were instantiated on a $20 \times 20 \times 20$ grid (nominal physical units).

We are particularly concerned with the order of accuracy of the method.
Decreasing the grid size will always increase the accuracy of the method, but with appropriate methods the accuracy improves to some power of the sampled spacing.
This is important for measuring morphometric quantities such as curvature that require a certain amount of smoothness in the embedding.

In general, the order of accuracy, $m$, is estimated by:
\begin{equation}
  m = \log_2 \left( \frac{\lVert \phi_h - \phi \rVert_p}{\lVert \phi_{h/2} - \phi  \rVert_p} \right)
\end{equation}
where $\phi$ is the ground truth, $\phi_h$ is the solution at spacing $h$, $\phi_{h/2}$ is the solution at spacing $h/2$, and $\lVert \cdot \rVert_p$ is an $\ell^p$ norm.
Since the distance transform is only Lipschitz continuous at the medial axis, different norms are used to approximate measures of the global (everywhere) and local (away from medial axis) order of accuracy.
Because the signed distance map is Lipschitz continuous, the method can only be first-order accurate when measured with the $\ell^\infty$ norm but can achieve arbitrary order of accuracy when measured with the $\ell^1$ norm~\cite{zhang2006high}.

\subsubsection{Order of Accuracy in the Narrowband}
\label{subsec:narrowband}
The order of accuracy for solving the narrowband is measured.
The analytic signed distance transform of the sphere, torus, and double-sphere are instantiated on a grid with a prescribed spacing.
Using Equation~\ref{eqn:density_construction}, phase densities $\rho_1 = 100$, $\rho_2 = 0$ (in arbitrary density units), and Heaviside smoothness constant $\epsilon = 2$, the density image is instantiated.
The narrowband is solved and the $\ell^1$ and $\ell^\infty$ norm measured on the narrowband voxels.
The grid size is divided by two and repeated, allowing an estimation of the order of accuracy of the narrowband initialization.
No smoothing is performed such to preserve the zero crossing location and the threshold is set to the average of the phase densities:
\begin{equation}
  T = \frac{\rho_1 + \rho_2}{2}
\end{equation}
One can see this threshold is correct from Equations~\ref{eqn:density_construction} and \ref{eqn:implicit_surface} noting that $\theta(0) = 0.5$.

\subsubsection{Order of Accuracy in Sweeping}
\label{subsec:sweeping}
The order of accuracy of the signed fast sweeping method is measured.
The analytic signed distance transform of the sphere, torus, and double-sphere are instantiated at the narrowband with a prescribed spacing.
The remainder of the domain is solved from the narrowband and the $\ell^1$ and $\ell^\infty$ norms are measured excluding narrowband voxels.
The grid size is divided by two and the process repeated, allowing an estimation of the order of accuracy of the extension method.

\subsubsection{Visualizing Result}
\label{subsec:visual}
The ideal signed distance transform of a double sphere is sampled at $h=0.25$ and mapped into density as in Section~\ref{subsec:narrowband}.
From the density image, the signed distance transform is solved using the narrowband and subsequent fast sweeping method.
Different level sets of the ideal and solved distance transform are visualized using the Marching Cubes algorithms~\cite{lorensen1987marching} to demonstrate the visual smoothness and correctness of the method.


\subsection{Trabecular Bone}
\label{subsection:experiments:trabecular_bone}
10 previously imaged bovine trabecular bone cubes were used to verify the correctness of the algorithm~\cite{sandino2013trabecular}.
Bone cubes were sawed to 10 mm in edge length and imaged at a nominal resolution of $\SI{20}{\micro\metre}$.
The bones were embedded with Equation~\ref{eqn:implicit_surface} using a standard deviation $\sigma = \SI{20}{\micro\metre}$ and threshold $T = 400~\text{mg HA/cc}$.
The signed distance transform was computed with the proposed method.

To increase the execution speed, the first-order fast sweeping method was used for extending the narrowband.
A high accuracy is only needed at the narrowband for visualization, phase density estimation, and morphometry.
The models here had an average of 143 million voxels, requiring roughly 20 minutes to compute the narrowband and roughly 40 minutes per iteration in the fast sweeping method on a consumer grade MacBook Pro (32 GB random access memory, 2.3 GHz processor with 8 cores).
The fast sweeping method can be accelerated by only solving to a certain distance, but a high-order cannot be parallelized due to the lack of a monotonicity condition~\cite{zhang2006high}.
It should be noted that the narrowband method is a so-called ``embarrassingly'' parallelizable algorithm, where every narrowband voxel could be solved on a separate thread.

\subsubsection{Embedding and Surface Localization}
A visual comparison is made between the density field, the implicit surface, and the solved signed distance transform to verify the location of the level set coincides between images.
The exact signed distance transform is also computed from the implicit surface to demonstrate the quantization artifact~\cite{maurer2003linear}.
Meshes are extracted using the Marching Cubes algorithms~\cite{lorensen1987marching} at the appropriate level set for the given field.

\subsubsection{Phase Densities}
Next, density estimation is validated.
The phase components are estimated using Equations~\ref{eqn:rho1} and \ref{eqn:rho2}.
Once estimated, the density image is reconstructed using Equation~\ref{eqn:density_construction} and compared visually.
The bone mineral density (vBMD [$\text{mg HA/cc}$]) between the original density image and the constructed density image are compared using regression and Bland-Altman analysis~\cite{bland1986statistical}.
The regularization parameter was set to $\epsilon = 2 h$.

\subsubsection{Morphometry}
Finally, morphometry of the implicit surface is validated.
Volume, area, average mean curvature, and Euler-Poincar\'e characteristic are measured directly from the embedding using previously developed methods~\cite{chan2001active,peng1999pde,besler2021morph}.
\begin{eqnarray}
  V &=& \int_\Omega \theta(-\phi) dV \\
  A &=& \int_\Omega \delta(-\phi) |\nabla \phi| dV \\
  \langle H \rangle &=& \frac{\int_\Omega H \delta(-\phi) |\nabla \phi| dV}{\int_\Omega dV} \\
  \langle K \rangle &=& \frac{\int_\Omega K \delta(-\phi) |\nabla \phi| dV}{\int_\Omega dV}  \\
  \bar{H} &=& \int_\Omega H \delta(-\phi) |\nabla \phi| dV 
\end{eqnarray}
\begin{eqnarray}
  \bar{K} &=& \int_\Omega K \delta(-\phi) |\nabla \phi| dV \\
  \chi &=& \frac{\bar{K}}{2\pi}
\end{eqnarray}
where $V$ is the object volume, $A$ is the object surface area, $H$ is the mean curvature, $K$ is the Gaussian curvature, $\bar{H}$, $\bar{K}$ are the total curvatures, $\langle H \rangle$, $\langle K \rangle$ are the surface average curvature, $\chi$ is the Euler-Poincar\'e characteristic, $\theta$ is the Heaviside function, and $\delta$ is the Dirac delta function.
The curvatures are estimated using fourth-order central differences~\cite{besler2021morph} and integration is performed using Simpson's rule from Section~\ref{subsec:component_estimation}.
Furthermore, the numerical estimation of $\theta$ in Equation~\ref{eqn:numerical_heaviside} is used.
Taking the derivative of Equation~\ref{eqn:numerical_heaviside} gives the numerical approximation to the Dirac delta function:
\begin{equation}
  \delta_\epsilon(x) = \begin{cases}
    \frac{1}{2\epsilon} [1 + \cos\left(\frac{\pi x}{\epsilon}\right)] & |x| \leq \epsilon \\
    0 & |x| < \epsilon 
  \end{cases}  
\end{equation}

From these definitions, traditional bone morphometrics~\cite{bouxsein2010guidelines} can be derived:
\begin{eqnarray}
  \text{BV} &=& \int_\Omega \theta(-\phi) dV \\
  \text{BS} &=& \int_\Omega \delta(\phi) |\nabla \phi| dV \\
  \text{SMI} &=& 12 \langle H \rangle \frac{\text{BV}}{\text{BS}} \\
  \text{TBPf} &=& 2 \langle H \rangle \\
  \text{Conn.D} &=& \frac{1 - \chi}{TV}
\end{eqnarray}
where BV [$\si{\milli\metre\cubed}$] is the bone volume, BS [$\si{\milli\metre\squared}$] is the bone surface area, SMI [-] is the structure model index~\cite{hildebrand1997quantification}, TBPf [$\si{\per\milli\metre}$] is the trabecular bone pattern factor~\cite{hahn1992trabecular}, Conn.D [$\si{\per\milli\metre\cubed}$] is the connectivity density~\cite{odgaard1993quantification}, and TV [$\si{\milli\meter\cubed}$] is the total volume of the image.
Bone volume fraction (BV/TV, [\%]) can be computed simply as the ratio of BV to TV.
The link between mean curvature and SMI was first described by Jinnai \text{et al.}~\cite{jinnai2002surface} and extended to TBPf by Stauber and M{\"u}ller~\cite{stauber2006volumetric}.

The proposed method is compared to the corresponding standard methods for each measure.
The original density image is blurred with the same Gaussian filter and thresheld to binarize the volume.
It is standard in bone morphometry to perform a largest connected component filter to force one bone component and one marrow component in the image.
In fact, connected component filtering is required to meet the assumptions on Betti numbers in connectivity density estimation~\cite{odgaard1993quantification}.
As a result, the standard methods are applied to the binary image after largest connected component filtering.
However, connected component filter cannot be performed in the proposed method in a straight forward manner.
As a result, Euler-Poincar\'e characteristic and not connectivity density is used for validation.
Additionally, since SMI and TBPf are both related to mean curvature, only SMI is used for comparison.
BV/TV, BS, SMI, and $\chi$ are compared between the proposed and standard methods using regression and Bland-Altman plots.
Standard analysis was performed with Image Processing Language (IPL v5.42, SCANCO Medical AG, Br{\"u}ttisellen, Switzerland).
Finally, curvature is visualized across the surface.
The Marching Cubes algorithm~\cite{lorensen1987marching} is run at the zero level set and the vertices of the mesh interpolated into the image to estimate local mean and Gaussian curvature.

\section{Results}
\label{sec:results}

\subsection{Correctness on Ideal Surfaces}
\label{subsection:results:ideal_surfaces}

\subsubsection{Order of Accuracy in the Narrowband}
Measured order of accuracy is reported in Table~\ref{table:narrowband}.
For the sphere and torus, the narrowband has an order of accuracy of 4 in both the $\ell^1$ and $\ell^\infty$ norms.
This is an order higher than expected from analysis, a result seen in similar studies~\cite{chopp2001some}.
Since the narrowband is smooth, the $\ell^\infty$ and $\ell^1$ norms have the same order of accuracy.

For sphere-sphere intersection, the $\ell^1$ order of accuracy is 2 while the $\ell^\infty$ order of accuracy is one.
The degenerate $\ell^1$ order of accuracy comes from the implicit surface itself not being smooth, so the order of accuracy of the method cannot be achieved.
Similarly, a degenerate $\ell^1$ order of accuracy is seen, reflecting the presence of a non-smooth solution in the narrowband.
The $\ell^\infty$ order of accuracy is worse than one at higher resolutions.
This is caused by points in the narrowband that are non-smooth being mapped to a wrong but collinear point on the surface.
At the sphere-sphere intersection, non-smooth locations are mapped onto one of the two spheres and not the intersection, while still satisfying the collinear requirement.

\begin{table}
  \centering
  \caption{Measured order of accuracy for the narrrowband method.}
  \label{table:narrowband}
  \begin{tabular}{llcccc}
    \hline
    & $h$ & $\ell^1$ & Order & $\ell^\infty$ & Order \\
    \hline
    \multirow{4}{*}{\begin{tabular}{@{}l@{}}Sphere\end{tabular}} &
    0.5 & $6.43\cdot 10^{-5}$ & - & $1.62\cdot 10^{-4}$ & -  \\
    &0.25 & $3.26\cdot 10^{-6}$ & 4.30 & $1.01\cdot 10^{-5}$ & 4.01 \\
    &0.125 & $1.91\cdot 10^{-7}$ & 4.09 & $5.77\cdot 10^{-7}$ & 4.12 \\
    &0.0625 & $1.17\cdot 10^{-8}$ & 4.02 & $3.80\cdot 10^{-8}$ & 3.92 \\
    \hline
    \multirow{4}{*}{\begin{tabular}{@{}l@{}}Torus\end{tabular}} &
    0.5 & $9.79\cdot 10^{-5}$ & - & $2.71\cdot 10^{-4}$ & -  \\
    &0.25 & $4.56\cdot 10^{-6}$ & 4.42 & $1.48\cdot 10^{-5}$ & 4.20 \\
    &0.125 & $2.64\cdot 10^{-7}$ & 4.11 & $8.80\cdot 10^{-7}$ & 4.07 \\
    &0.0625 & $1.60\cdot 10^{-8}$ & 4.04 & $5.52\cdot 10^{-8}$ & 3.99 \\
    \hline
    \multirow{4}{*}{\begin{tabular}{@{}l@{}}Double\\Spheres\end{tabular}} &
    0.5 & $1.33\cdot 10^{-3}$ & - & $1.27\cdot 10^{-1}$ & -  \\
    &0.25 & $3.00\cdot 10^{-4}$ & 2.15 & $8.68\cdot 10^{-2}$ & 0.54 \\
    &0.125 & $7.20\cdot 10^{-5}$ & 2.06 & $3.88\cdot 10^{-2}$ & 1.16 \\
    &0.0625 & $1.80\cdot 10^{-5}$ & 2.00 & $3.52\cdot 10^{-2}$ & 0.14 \\
    \hline
   \end{tabular}
\end{table}

\subsubsection{Order of Accuracy in Sweeping}
Measured order of accuracy in sweeping is reported in Table~\ref{table:extension}.
For the sphere and torus, an order of accuracy of $3$ is seen in the $\ell^1$ norm with an order of accuracy of $1$ in the $\ell^\infty$ norm.
The reduced $\ell^\infty$ corresponds to a shock at their center.
Across all grid sizes, the number of iterations was less than 40

The non-smooth sphere-sphere intersection demonstrated an order of accuracy near 2 for the $\ell^1$ norm and an order of accuracy of $1$ in the $\ell^\infty$ norm.
Errors were seen in areas of rarefaction, corresponding to the lens shape of the intersecting spheres, degrading the $\ell^1$ norm order of accuracy.
Outside the area of rarefaction the result was as expected.
This result is well known in the sweeping method~\cite{zhang2006high}.
The number of iterations required for convergence was larger for the non-smooth surface.

\begin{table}
  \centering
  \caption{Measured order of accuracy for the fast sweeping method.}
  \label{table:extension}
  \begin{tabular}{@{}llccccc@{}}
    \hline
    & $h$ & $\ell^1$ & Order & $\ell^\infty$ & Order & Iter. \\
    \hline
    \multirow{4}{*}{\begin{tabular}{@{}l@{}}Sphere\end{tabular}} &
    0.5 & $1.20\cdot 10^{-3}$ & - & $1.81\cdot 10^{-1}$ & - & 17 \\
    &0.25 & $1.78\cdot 10^{-4}$ & 2.76 & $9.14\cdot 10^{-2}$ & 0.99 & 20 \\
    &0.125 & $1.49\cdot 10^{-5}$ & 3.58 & $4.56\cdot 10^{-2}$ & 1.00 & 23 \\
    &0.0625 & $3.78\cdot 10^{-7}$ & 5.30  & $2.27\cdot 10^{-2}$ & 1.01 & 24 \\
    \hline
    \multirow{4}{*}{\begin{tabular}{@{}l@{}}Torus\end{tabular}} &
    0.5 & $3.22\cdot 10^{-3}$ & - & $1.10\cdot 10^{-1}$ & - & 24 \\
    &0.25 & $6.07\cdot 10^{-4}$ & 2.40 & $6.01\cdot 10^{-2}$ & 0.87 & 18 \\
    &0.125 & $8.13\cdot 10^{-5}$ & 2.90 & $3.05\cdot 10^{-2}$ & 0.98 & 23 \\
    &0.0625 & $4.98\cdot 10^{-6}$ & 4.03  & $1.52\cdot 10^{-2}$ & 1.01 & 35 \\
    \hline
    \multirow{4}{*}{\begin{tabular}{@{}l@{}}Double\\Spheres\end{tabular}} &
    0.5 & $3.42\cdot 10^{-3}$ & - & $1.39\cdot 10^{-1}$ & - & 42 \\
    &0.25 & $9.02\cdot 10^{-4}$ & 1.92 & $9.30\cdot 10^{-2}$ & 0.58 & 41 \\
    &0.125 & $3.07\cdot 10^{-4}$ & 1.56 & $4.62\cdot 10^{-2}$ & 1.01 & 25 \\
    &0.0625 & $1.21\cdot 10^{-4}$ & 1.34  & $2.28\cdot 10^{-2}$ & 1.02 & 31 \\
    \hline
   \end{tabular}
\end{table}

\begin{figure}[!h]
  \centering
  \begin{tabular}{cc}
    \subfloat[$\phi = 3$]{
      \includegraphics[width=0.33\linewidth]{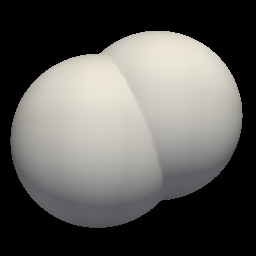}%
      \label{fig:mc:ideal:p3}
    } &
    \subfloat[$\tilde{\phi} = 3$]{
      \includegraphics[width=0.33\linewidth]{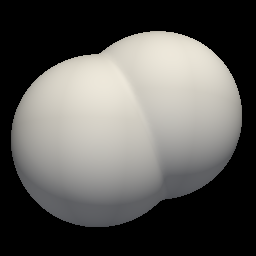}%
      \label{fig:mc:solved:p3}
    } \\
    \subfloat[$\phi = 0$]{
      \includegraphics[width=0.33\linewidth]{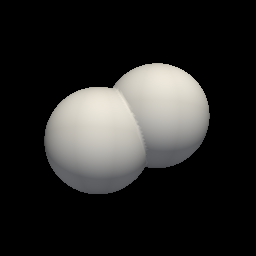}%
      \label{fig:mc:ideal:p0}
    } &
    \subfloat[$\tilde{\phi} = 0$]{
      \includegraphics[width=0.33\linewidth]{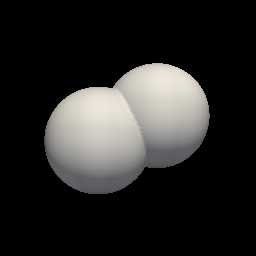}%
      \label{fig:mc:solved:p0}
    } \\
    \subfloat[$\phi = -3$]{
      \includegraphics[width=0.33\linewidth]{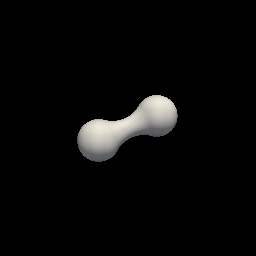}%
      \label{fig:mc:ideal:n3}
    } &
    \subfloat[$\tilde{\phi} = -3$]{
      \includegraphics[width=0.33\linewidth]{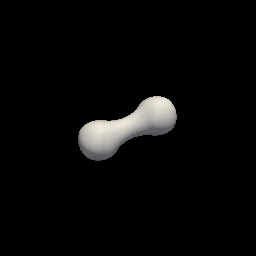}%
      \label{fig:mc:solved:n3}
    } \\
    \subfloat[$\phi = -4$]{
      \includegraphics[width=0.33\linewidth]{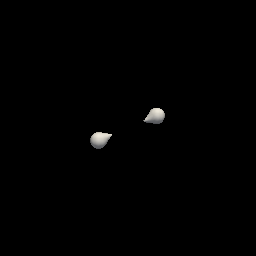}%
      \label{fig:mc:ideal:n4}
    } &
    \subfloat[$\tilde{\phi} = -4$]{
      \includegraphics[width=0.33\linewidth]{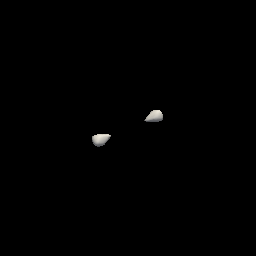}%
      \label{fig:mc:solved:n4}
    }
  \end{tabular}
  \caption{Visualization of the ideal ($\phi$) and solved ($\tilde{\phi}$) double sphere surface for $h=0.25$. Level set values for visualization are given below the figures.}
  \label{fig:mc}
\end{figure}

\begin{figure*}
  \centering
  \begin{tabular}{ccc}
    \subfloat[Contours of $\rho$ and $\psi$]{
      \includegraphics[width=0.3\linewidth]{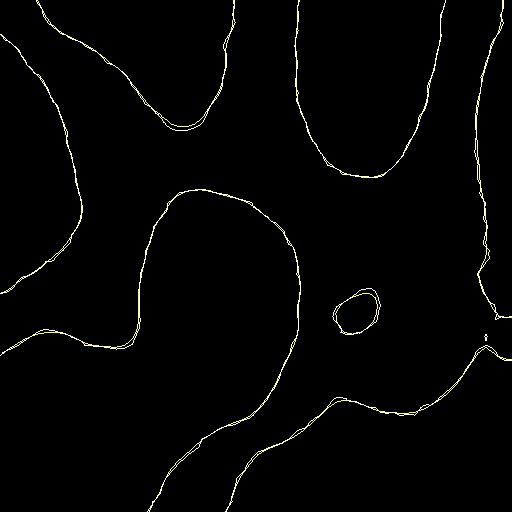}%
      \label{fig:outline:rho}
    } &
    \subfloat[Contours of $\phi$ and $\psi$]{
      \includegraphics[width=0.3\linewidth]{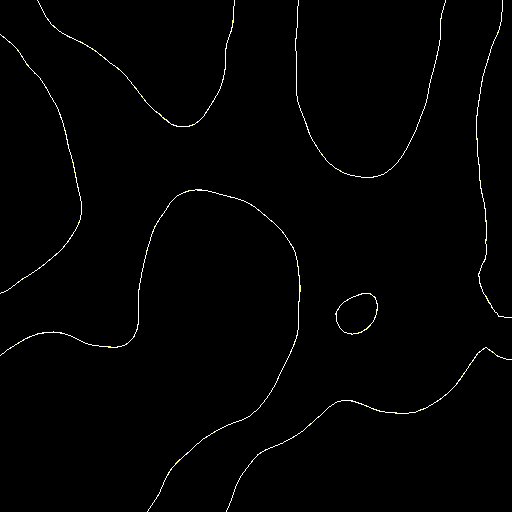}%
      \label{fig:outline:hsdt}
    } &
    \subfloat[Contours of Exact SDT and $\psi$]{
      \includegraphics[width=0.3\linewidth]{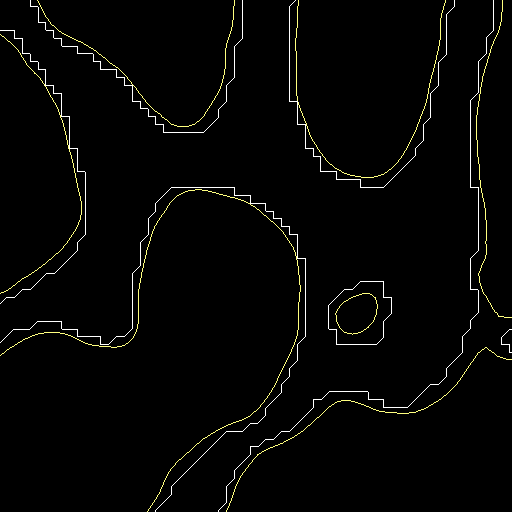}%
      \label{fig:outline:exact}
    }
  \end{tabular}
  \caption{Visualizing the level sets of the embeddings. (\ref{fig:outline:rho}) $\rho$ is outlined in white and $\psi$ is outlined in yellow. Blurring removed noise and smoothed the surface without modifying the location severely. (\ref{fig:outline:hsdt}) $\phi$ is outlined in white and $\psi$ is outlined in yellow.  The signed distance transforms tracks $\psi$ almost perfectly. (\ref{fig:outline:hsdt}) The exact signed distance transform is outlined in white and $\psi$ is outlined in yellow. The quantization artifact can be seen as square edges.}
  \label{fig:outline}
\end{figure*}

\begin{figure*}
  \centering
  \begin{tabular}{cc}
    \subfloat[$\rho$]{
      \includegraphics[width=0.39\linewidth]{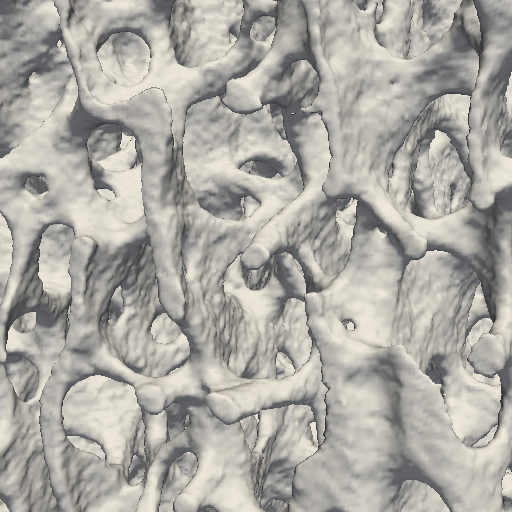}%
      \label{fig:surface:rho}
    } &
    \subfloat[$\psi = T - G_\sigma * \rho$]{
      \includegraphics[width=0.39\linewidth]{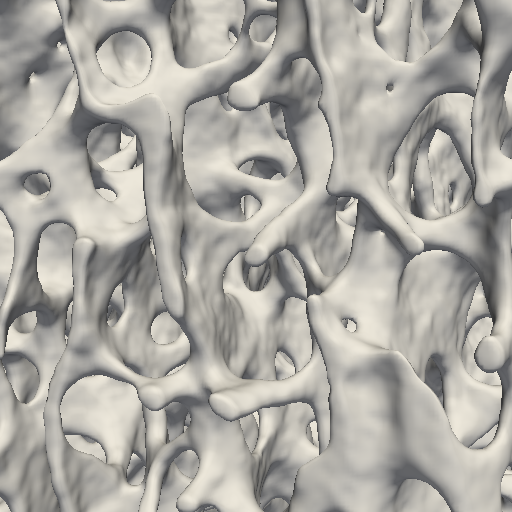}%
      \label{fig:surface:psi}
    } \\
    \subfloat[$\phi$]{
      \includegraphics[width=0.39\linewidth]{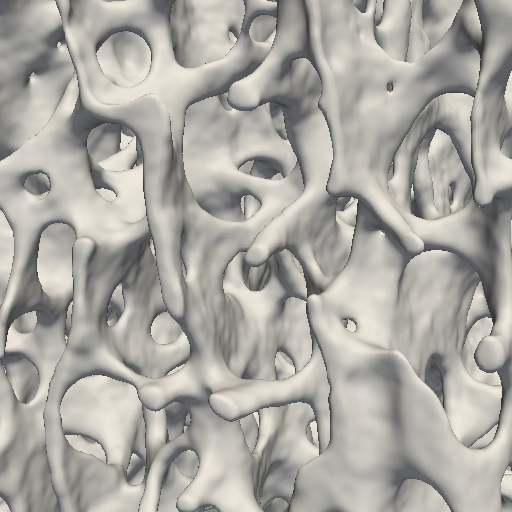}%
      \label{fig:surface:hsdt}
    } &
    \subfloat[Exact SDT of $\psi$]{
      \includegraphics[width=0.39\linewidth]{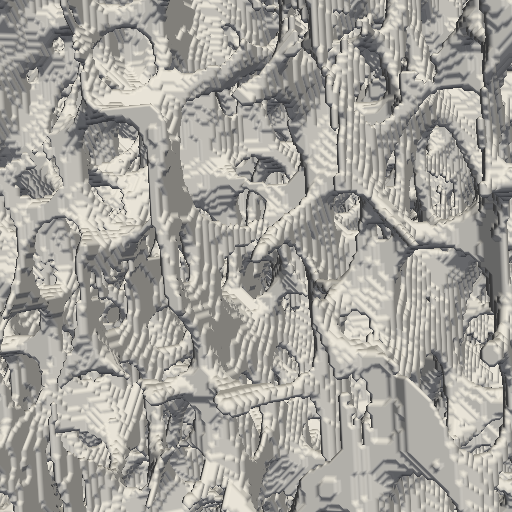}%
      \label{fig:surface:exact}
    }
  \end{tabular}
  \caption{Visualizing the surfaces of the trabecular bone. (\ref{fig:surface:rho}) The native density image exhibits noise which is removed after filtering (\ref{fig:surface:psi}). (\ref{fig:surface:hsdt}) The high-order signed distance transform agrees with the location and smoothness of $\psi$. (\ref{fig:surface:exact}) The exact signed distance transform of $\psi$ exhibits quantization artifacts.}
  \label{fig:surface}
\end{figure*}

\subsubsection{Visualizing Result}
\label{subsec:result:visual}
The ideal and solved double sphere surface is rendered at four level sets in Figure~\ref{fig:mc}.
The zero level set surfaces correspond very well with no visual indication of the quantization artifact.
In the positive domain, the level set corresponding to a distance $3$ from the surface shows a slight smoothing of the intersection, corresponding with the ability to localize shocks.
In the negative domain, errors in the thickness of the dumbbell shape can be seen, corresponding to errors in rarefaction.
Finally, the dumbbell separates into two tear-drop shapes, some errors can be seen in the roundness of the solution.


\subsection{Trabecular Bone}
\label{subsection:results:trabecular_bone}

\subsubsection{Embedding and Surface Localization}
The surface contour of $\rho$, $\psi$, and $\phi$ for a single slice are overlaid in Figure~\ref{fig:outline}.
Gaussian blurring reduces noise and smooths the surface while slightly modifying the level set location.
However, no further modification to the level set is seen in the solved signed distance transform.
The exact signed distance transform produces large errors as expected.
The meshed level sets are displayed in Figure~\ref{fig:surface}.
Noise has been removed by blurring and the solved distance transform agrees with the filtered image.
Importantly, the exact signed distance transform applied to the binarized image exhibits artifacts due to quantization of the metric.

\subsubsection{Phase Densities}
Measured phase densities are reported in Table~\ref{table:phase_densities}.
The average bone phase density is $646.6	\pm 34.0~\text{mg HA/cc}$ ($3734.8	\pm 192.0~\text{HU}$) and the average marrow phase density is $3.8	\pm 26.5~\text{mg HA/cc}$ ($100.9	\pm 149.8~\text{HU}$), reported $\text{mean} \pm \text{standard deviation}$.
Average densities are separated on either side of the $400~\text{mg HA/cc}$ threshold.
Negative effective densities are seen in the marrow space because the calibration model does not account for densities below water.
However, the marrow space Hounsfield units are in appropriate ranges between fat ($-100~\text{HU}$) and general soft tissue ($100~\text{HU}$).
The Hounsfield units of the bone phase are very large, but are reasonable given the high resolution of the imager.
When weighting the bone and marrow components by their volume fraction (Section~\ref{subsubsec:morphometry}), the image Hounsfield units range from $535.5~\text{HU}$ to $1575.9~\text{HU}$, consistent with clinical computed tomography where partial voluming would average the components together.
Consistency in the phase densities is expected since these were the same species at the same location and there were no known diseases affecting mineralization or marrow composition.

\begin{table}
  \centering
  \caption{Measured phase densities.}
  \label{table:phase_densities}
  \begin{tabular}{ccccc}
    \hline
    Sample & $\rho_1~\text{[mg HA/cc]}$ & $\rho_2~\text{[mg HA/cc]}$ & $\rho_1~\text{[HU]}$ & $\rho_2~\text{[HU]}$ \\
    \hline
    1 & 628.0	& -33.3 & 3629.5 & -109.1 \\
    2 & 720.3	& 5.5 & 4151.2 & 110.4 \\
    3 & 645.2	& 11.1 & 3726.5 & 142.2 \\
    4 & 652.1	& 42.0 & 3765.9 & 316.5 \\
    5 & 650.3	& 26.3 & 3755.4 & 227.9 \\
    6 & 643.5	& -5.1 & 3717.1 & 50.6 \\
    7 & 681.0	& 5.9 & 3929.2 & 112.4 \\
    8 & 620.0	& -39.5 & 3584.5 & -144.1 \\
    9 & 598.9	& -7.4 & 3465.0 & 37.2 \\
    10 & 626.9 & 32.8 & 3623.4 & 264.7 \\
    \hline
   \end{tabular}
\end{table}

\begin{figure}
  \centering
  \begin{tabular}{cc}
    \subfloat[Ground Truth]{
      \includegraphics[width=0.43\linewidth]{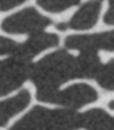}%
      \label{fig:density:psi}
    } &
    \subfloat[Equation~\ref{eqn:density_construction}]{
      \includegraphics[width=0.43\linewidth]{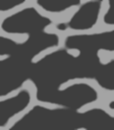}%
      \label{fig:density:constructed}
    }
  \end{tabular}
  \caption{(\ref{fig:density:psi}) Comparison of the filtered density image ($\psi$) and (\ref{fig:density:constructed}) that constructed from the high-order signed distance transform and estimated phase densities. Images are displayed in the same dynamic range.}
  \label{fig:density}
\end{figure}

\begin{figure}
  \centering
  \begin{tabular}{c}
    \subfloat[Regression]{
      \includegraphics[width=0.9\linewidth]{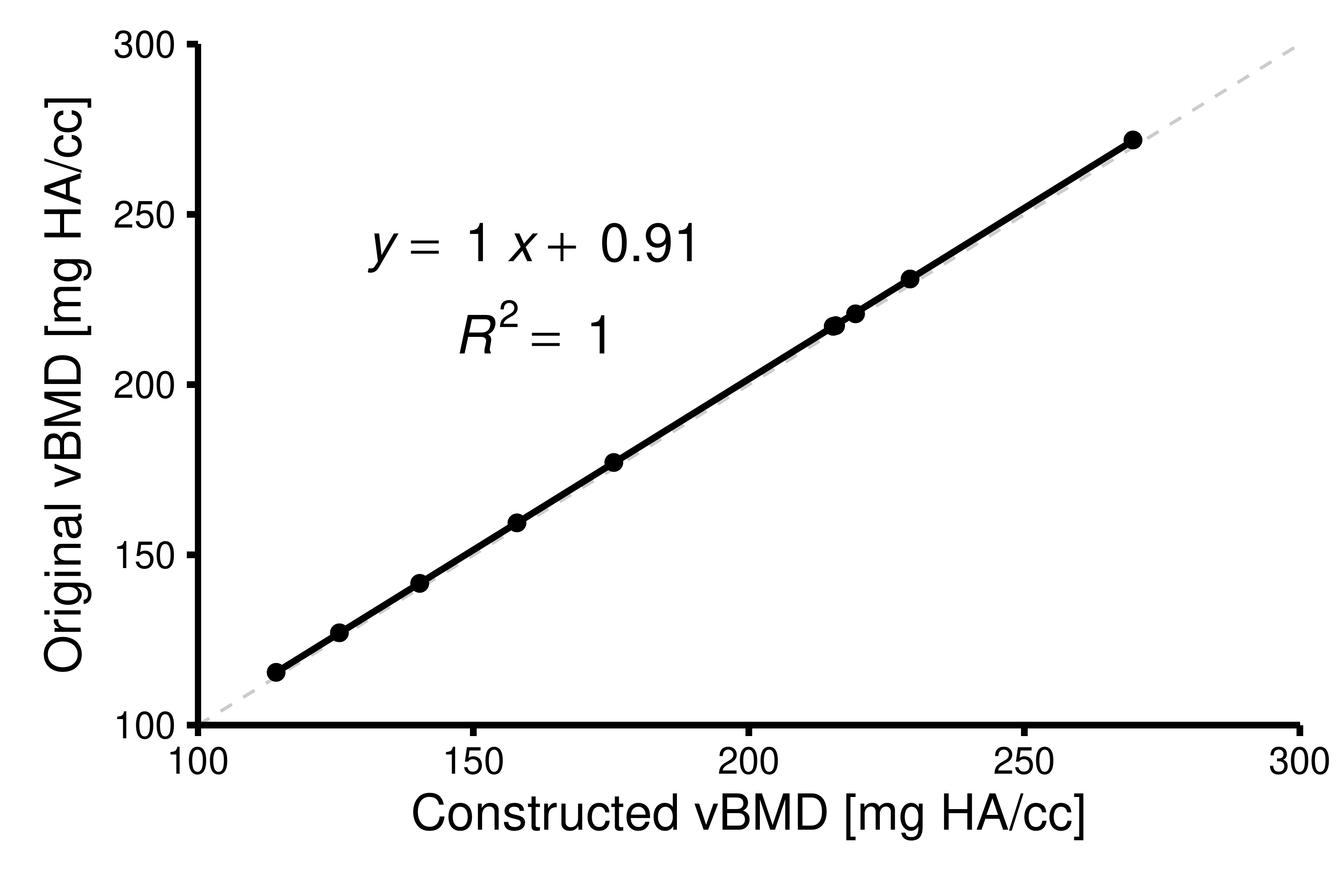}%
      \label{fig:agreement:regression}
    } \\
    \subfloat[Bland-Altman]{
      \includegraphics[width=0.9\linewidth]{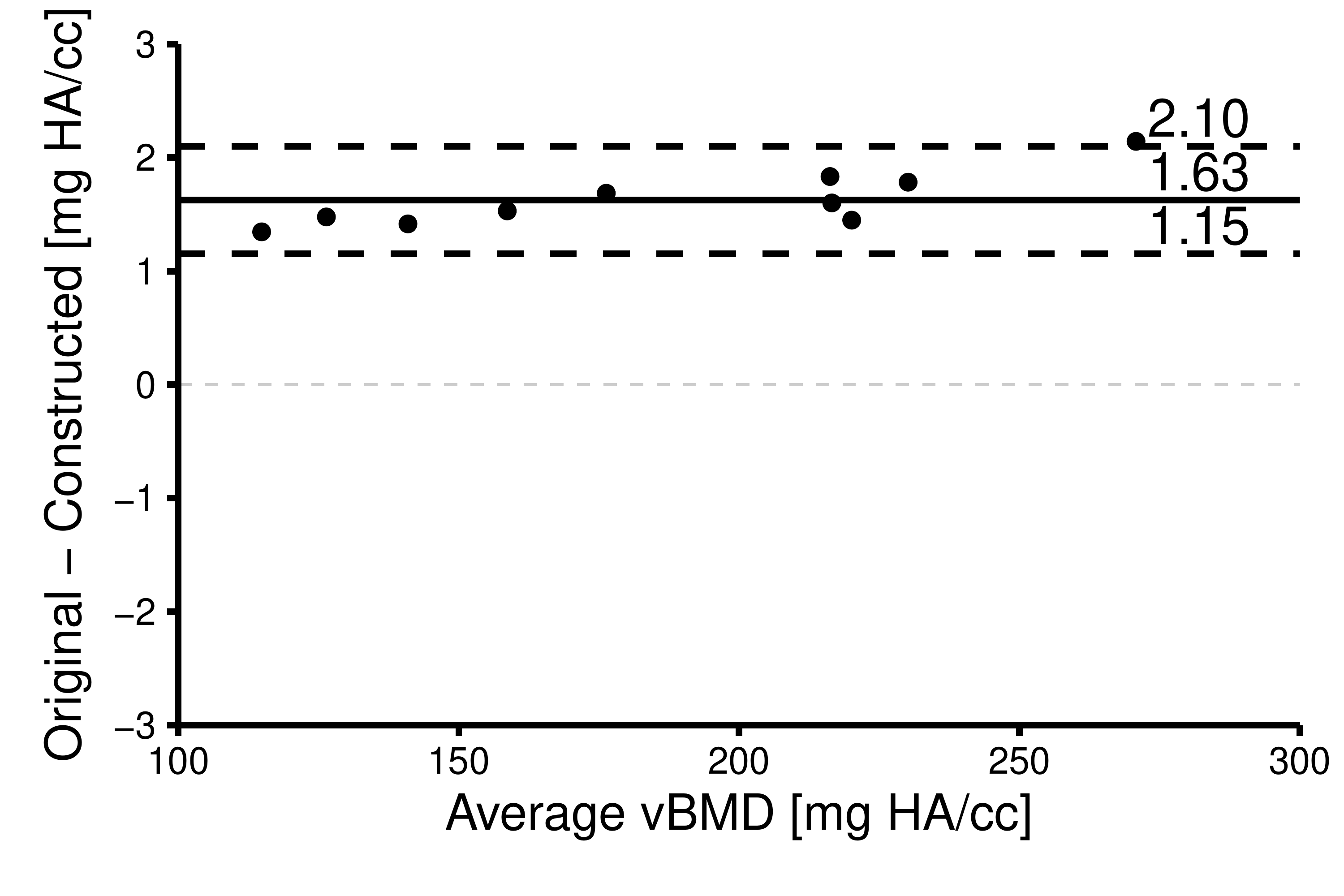}%
      \label{fig:agreement:ba}
    }
  \end{tabular}
  \caption{(\ref{fig:agreement:regression}) Regression and (\ref{fig:agreement:ba}) Bland-Altman analysis between the vBMD [mg HA/cc] of the smoothed density image $G_\sigma * \rho$ and the constructed density image of Equation~\ref{eqn:density_construction}.}
  \label{fig:agreement}
\end{figure}

\begin{figure*}
  \centering
  \begin{tabular}{cc}
    \subfloat[Regression BV/TV]{
      \includegraphics[width=0.44\linewidth]{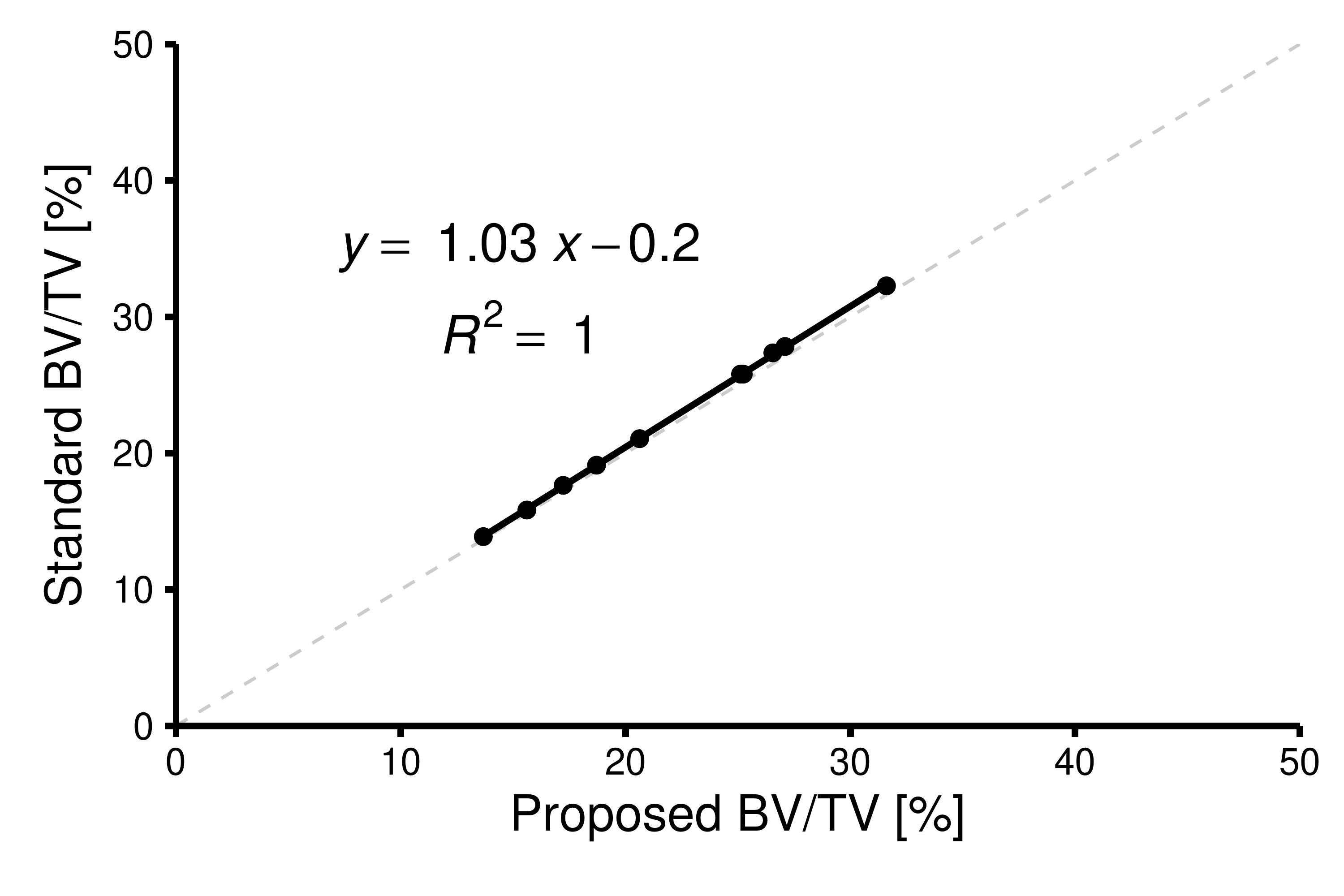}%
      \label{fig:morph:regression:bvtv}
    } &
    \subfloat[Bland-Altman BV/TV]{
      \includegraphics[width=0.44\linewidth]{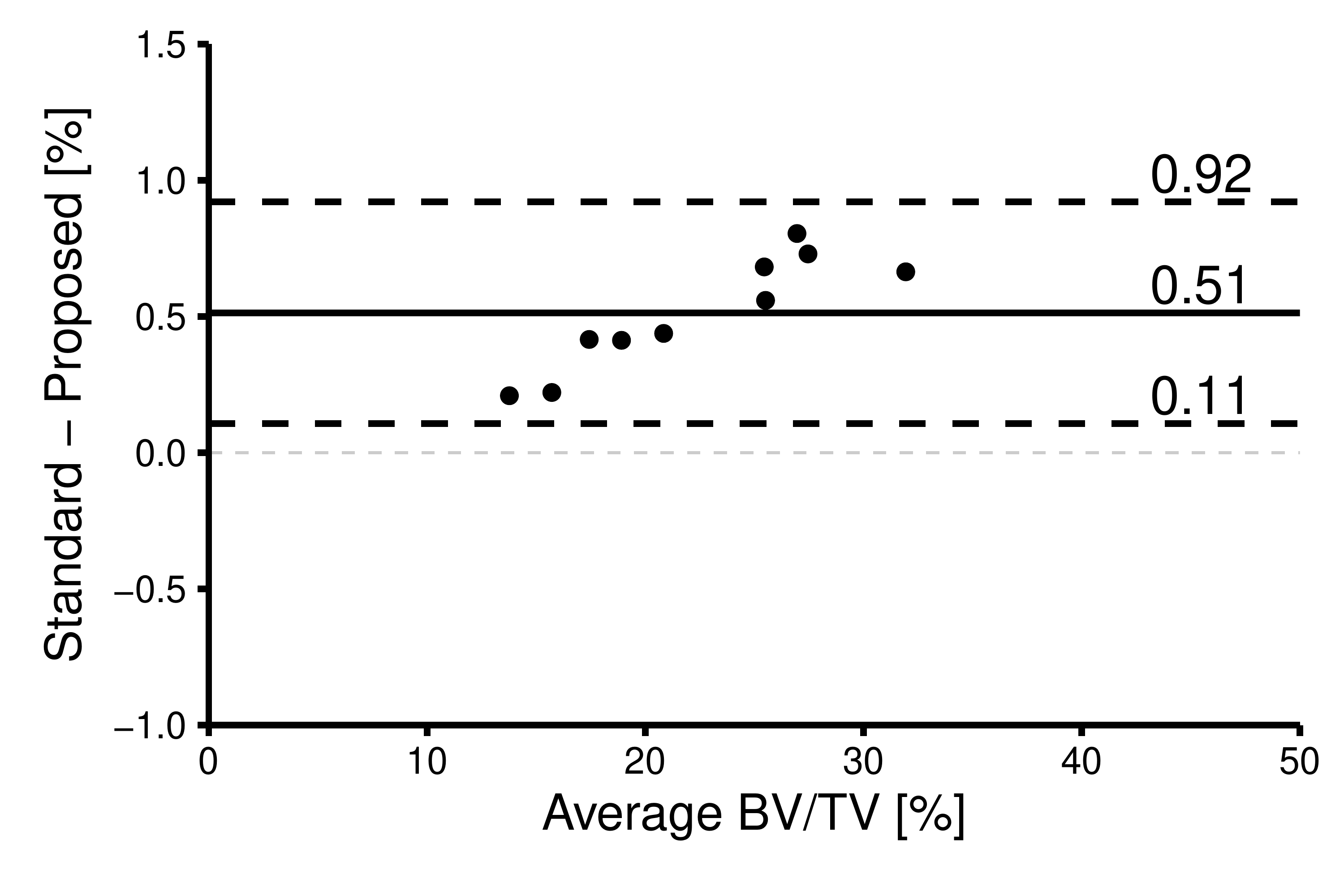}%
      \label{fig:morph:ba:bvtv}
    } \\
    \subfloat[Regression BS]{
      \includegraphics[width=0.44\linewidth]{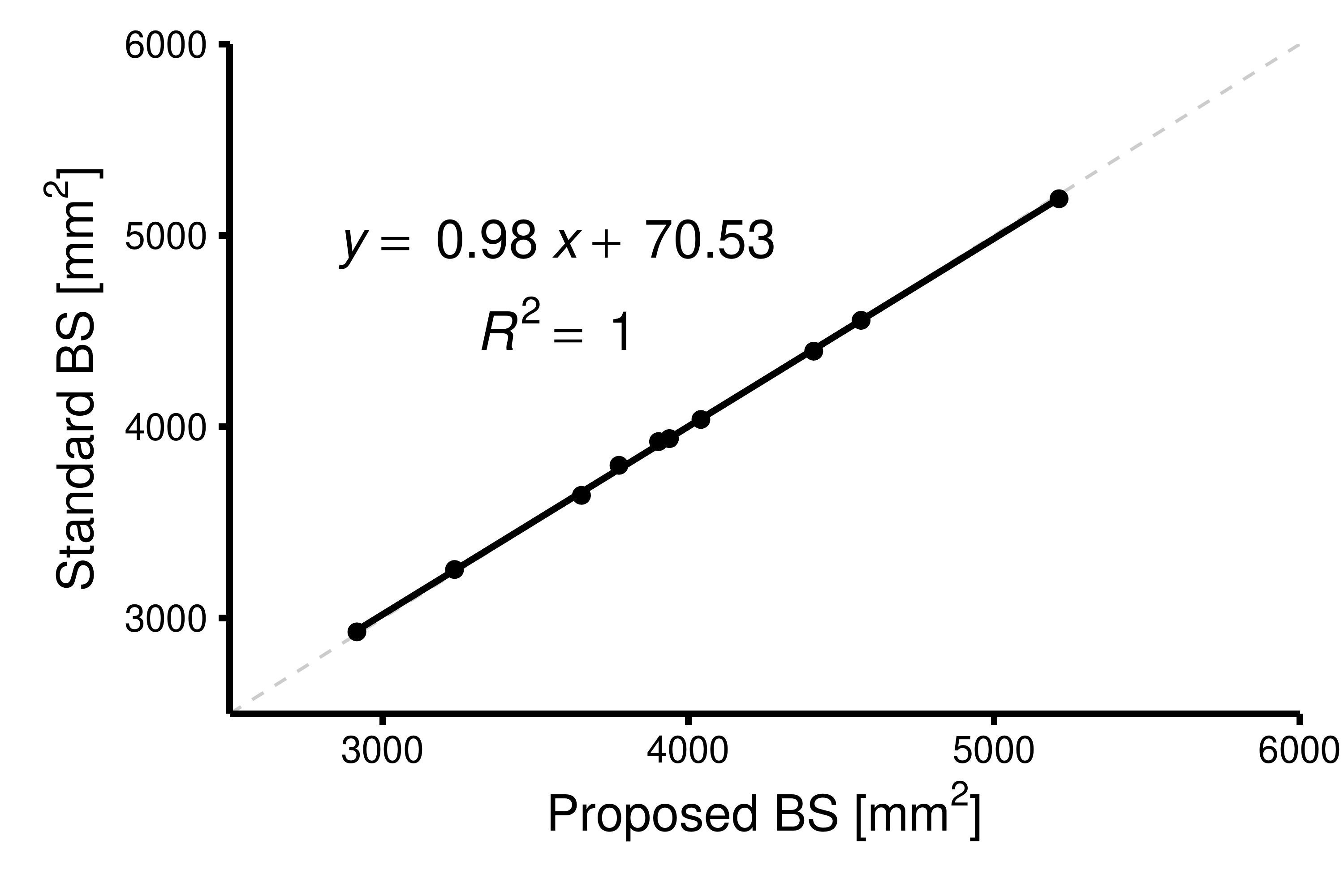}%
      \label{fig:morph:regression:bs}
    } &
    \subfloat[Bland-Altman BS]{
      \includegraphics[width=0.44\linewidth]{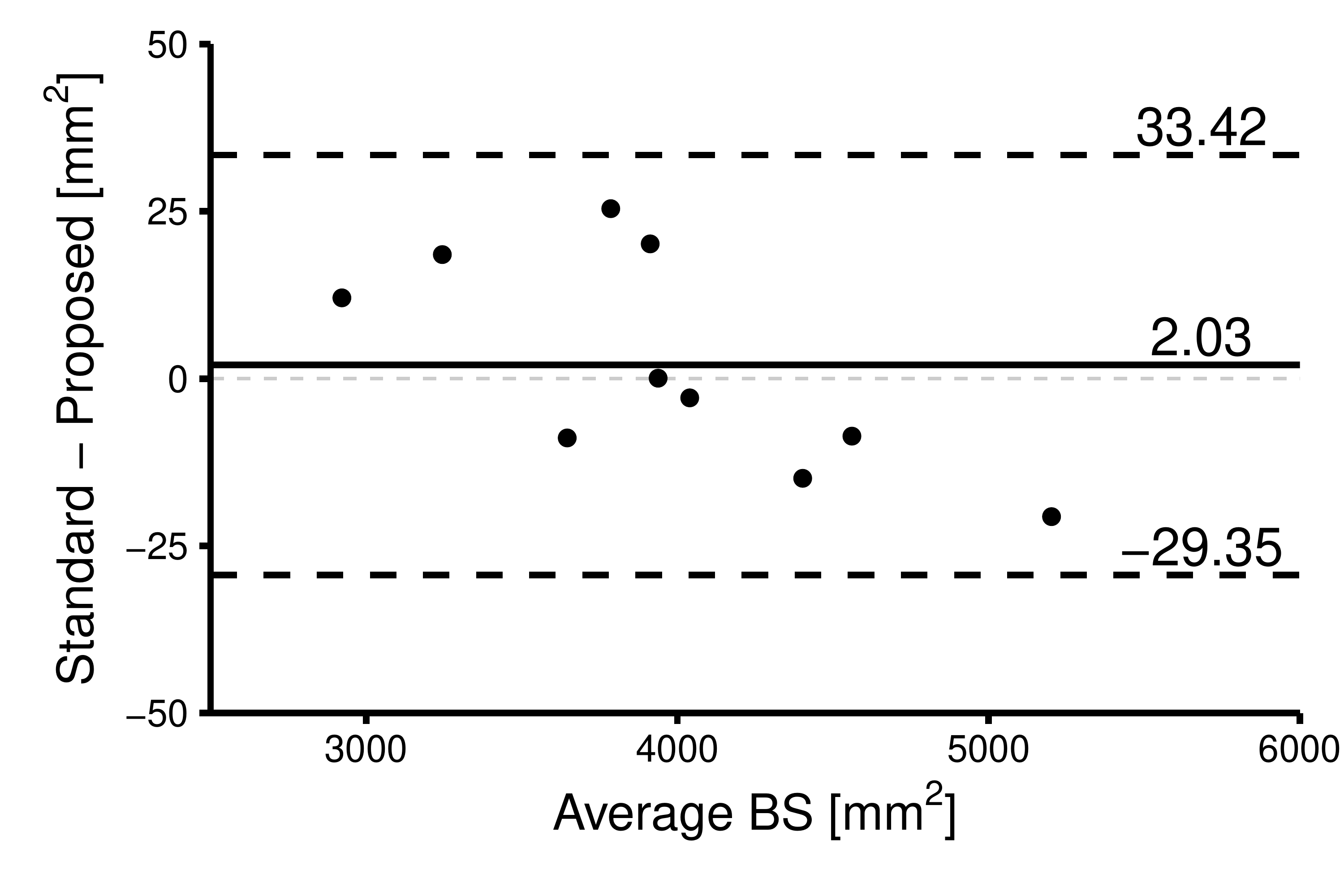}%
      \label{fig:morph:ba:bs}
    } \\
    \subfloat[Regression SMI]{
      \includegraphics[width=0.44\linewidth]{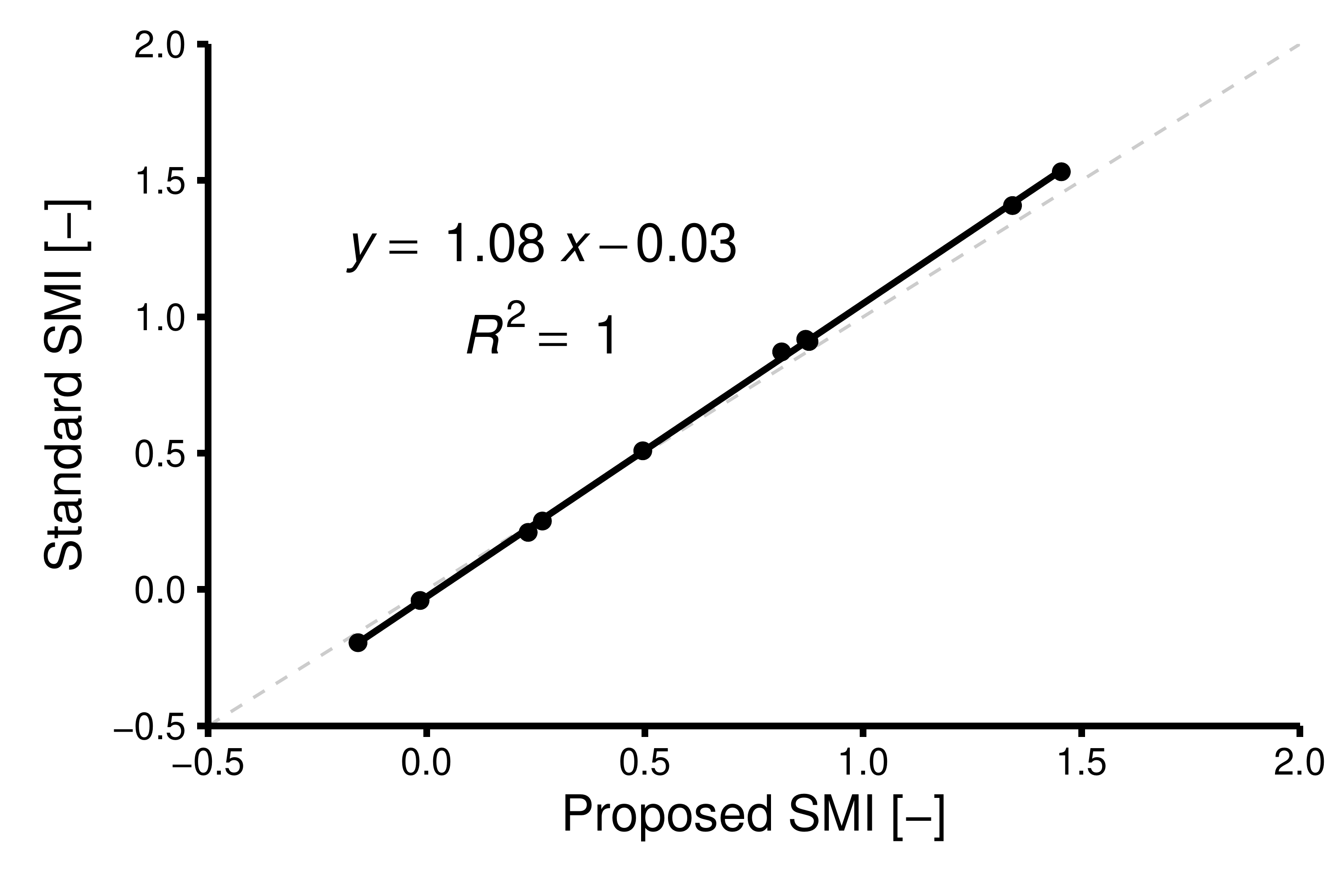}%
      \label{fig:morph:regression:smi}
    } &
    \subfloat[Bland-Altman SMI]{
      \includegraphics[width=0.44\linewidth]{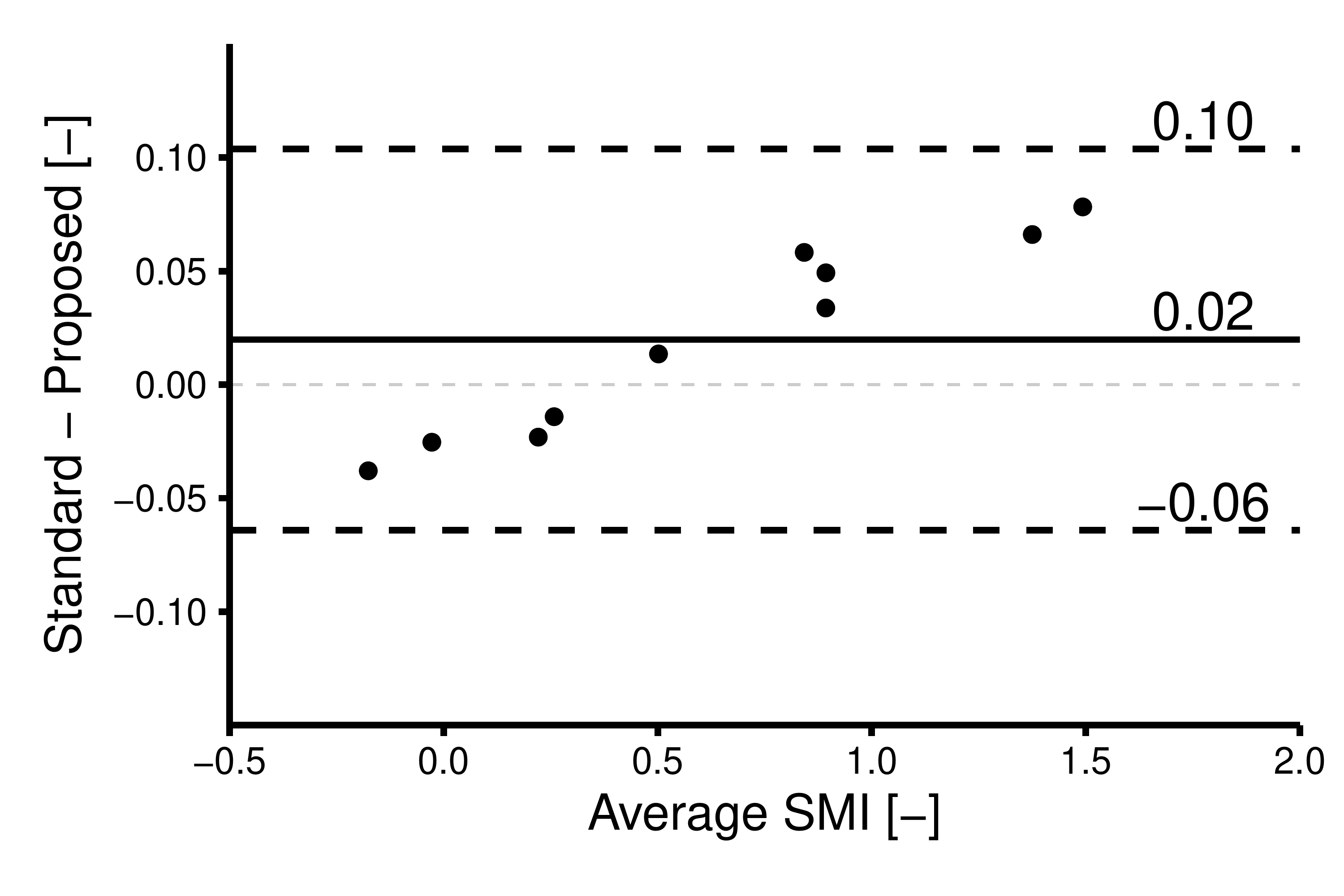}%
      \label{fig:morph:ba:smi}
    } \\
    \subfloat[Regression $\chi$]{
      \includegraphics[width=0.44\linewidth]{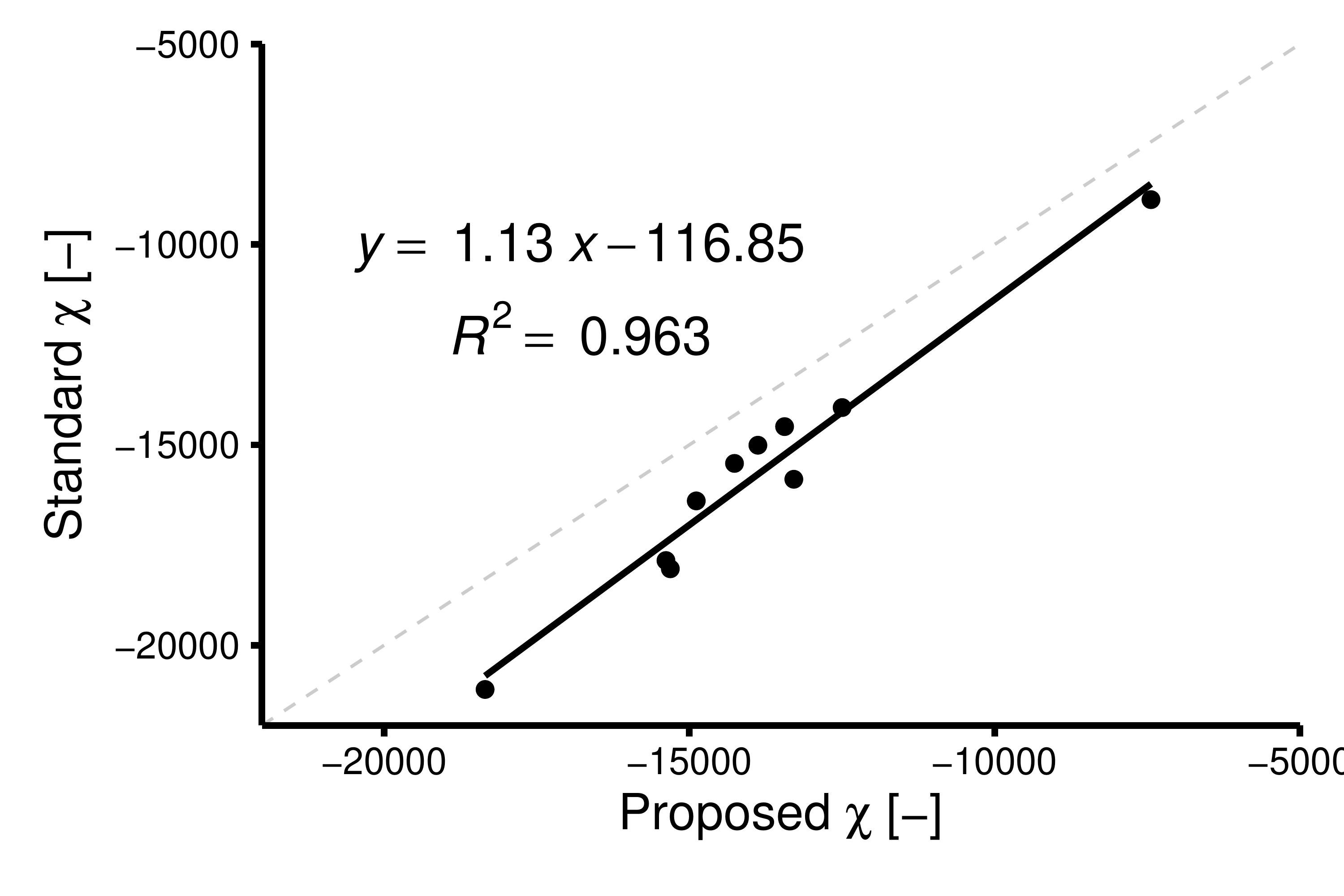}%
      \label{fig:morph:regression:chi}
    } &
    \subfloat[Bland-Altman $\chi$]{
      \includegraphics[width=0.44\linewidth]{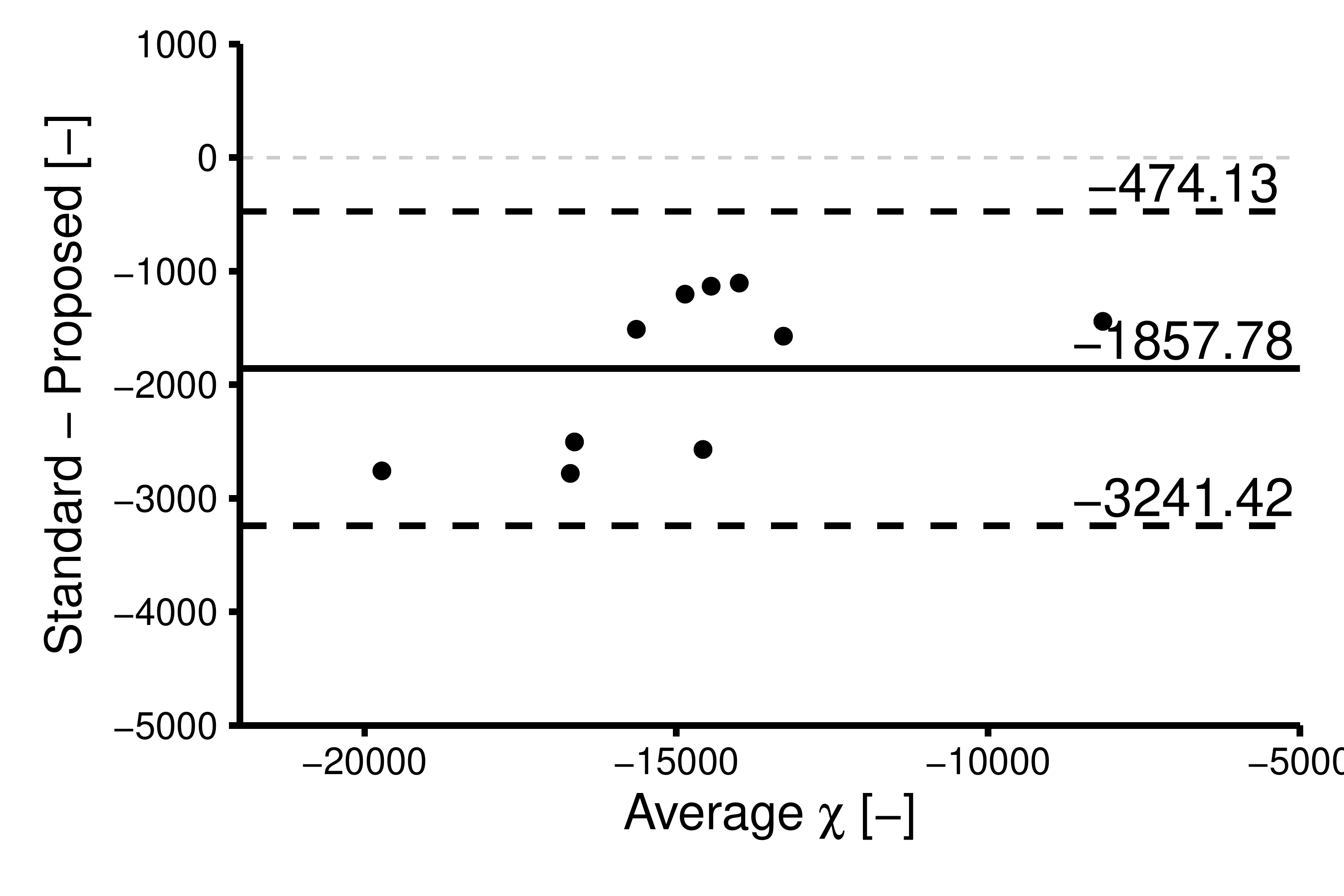}%
      \label{fig:morph:ba:chi}
    }
  \end{tabular}
  \caption{Regression and Bland-Altman plots between the proposed and standard method for bone volume fraction, bone surface area, Euler-Poincar\'e characteristic, and structure model index. Differences are a result of component labelling and methodology. An ideal relationship is shown in the dashed, light gray line.}
  \label{fig:morph}
\end{figure*}

\begin{figure*}[b]
  \centering
  \begin{tabular}{ccc}
    \multirow[c]{2}{0.35\linewidth}[105pt]{
      \subfloat[Mean Curvature]{
        \includegraphics[width=\linewidth]{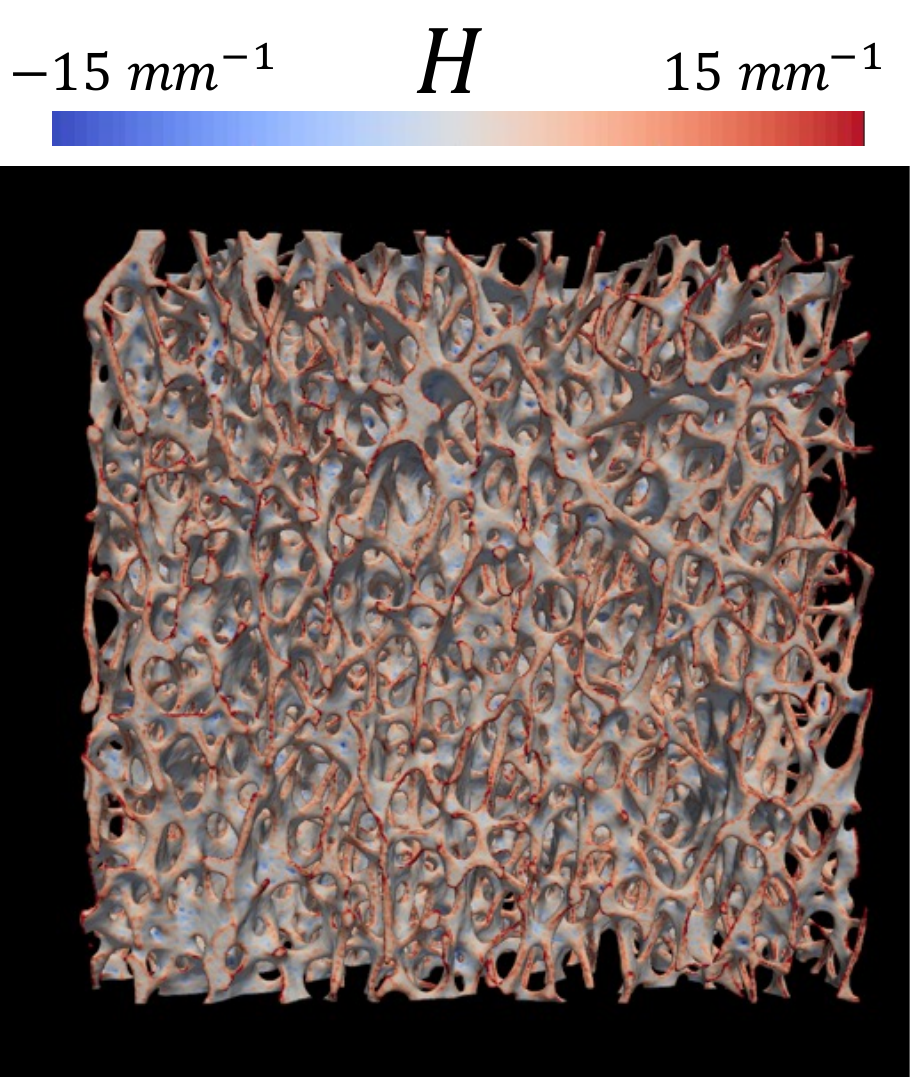}%
        \label{fig:curvature:h_surf}
      }
    } &
    \multirow[c]{2}{0.35\linewidth}[105pt]{
      \subfloat[Gaussian Curvature]{
        \includegraphics[width=\linewidth]{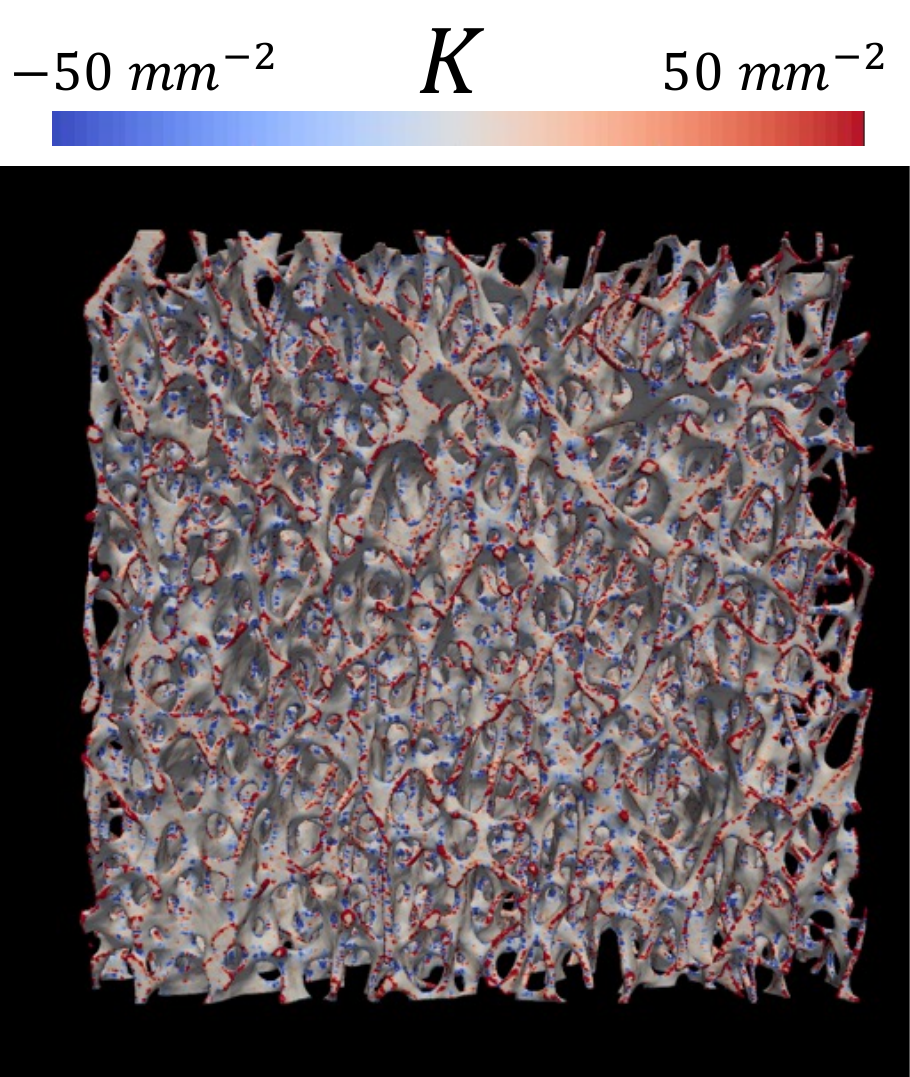}%
        \label{fig:curvature:k_surf}
      }
    } &
    \subfloat[$H$ Histogram]{
      \includegraphics[width=0.2\linewidth]{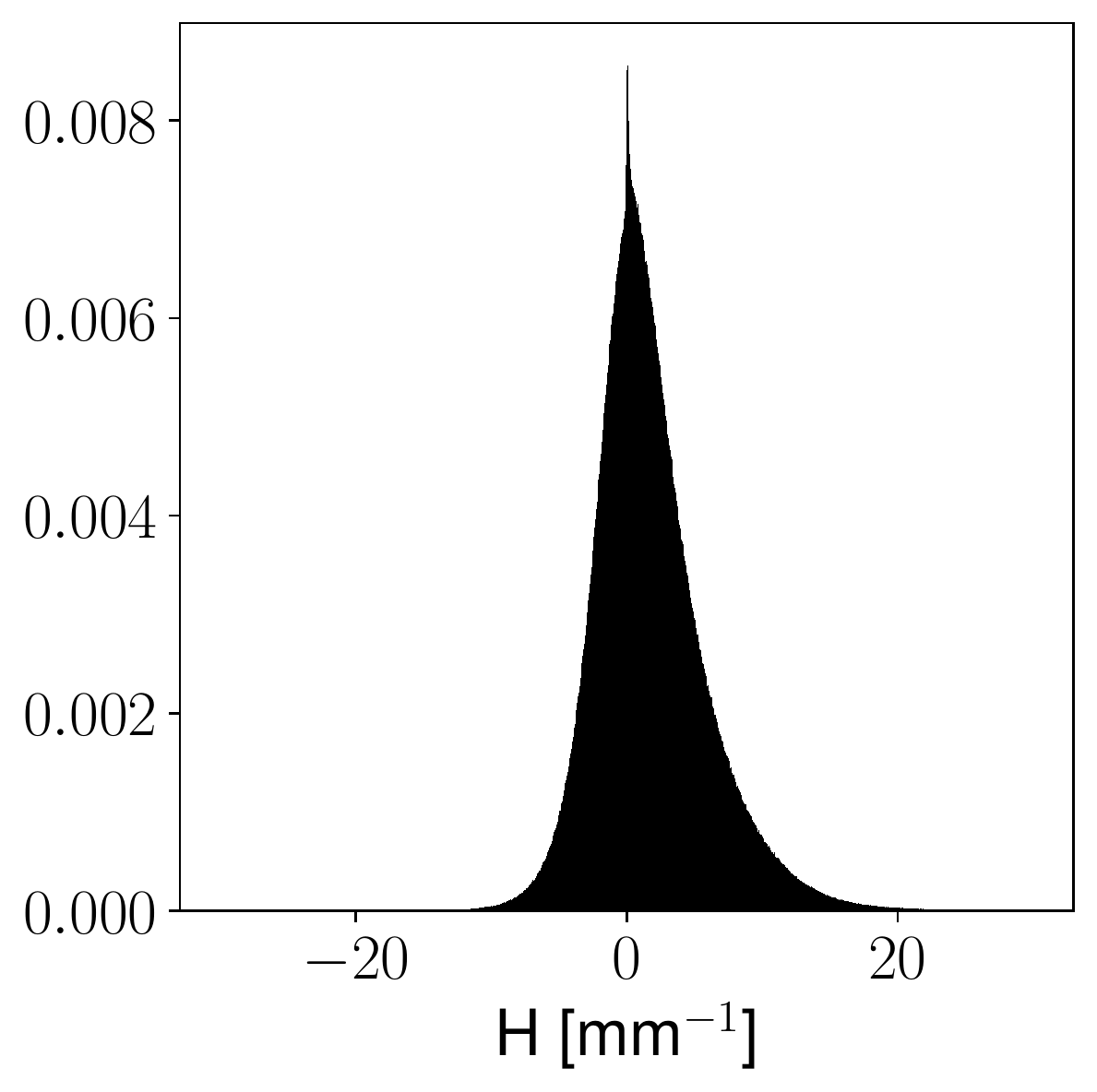}%
      \label{fig:curvature:h_hist}
    } \\
    & &
    \subfloat[$K$ Histogram]{
      \includegraphics[width=0.2\linewidth]{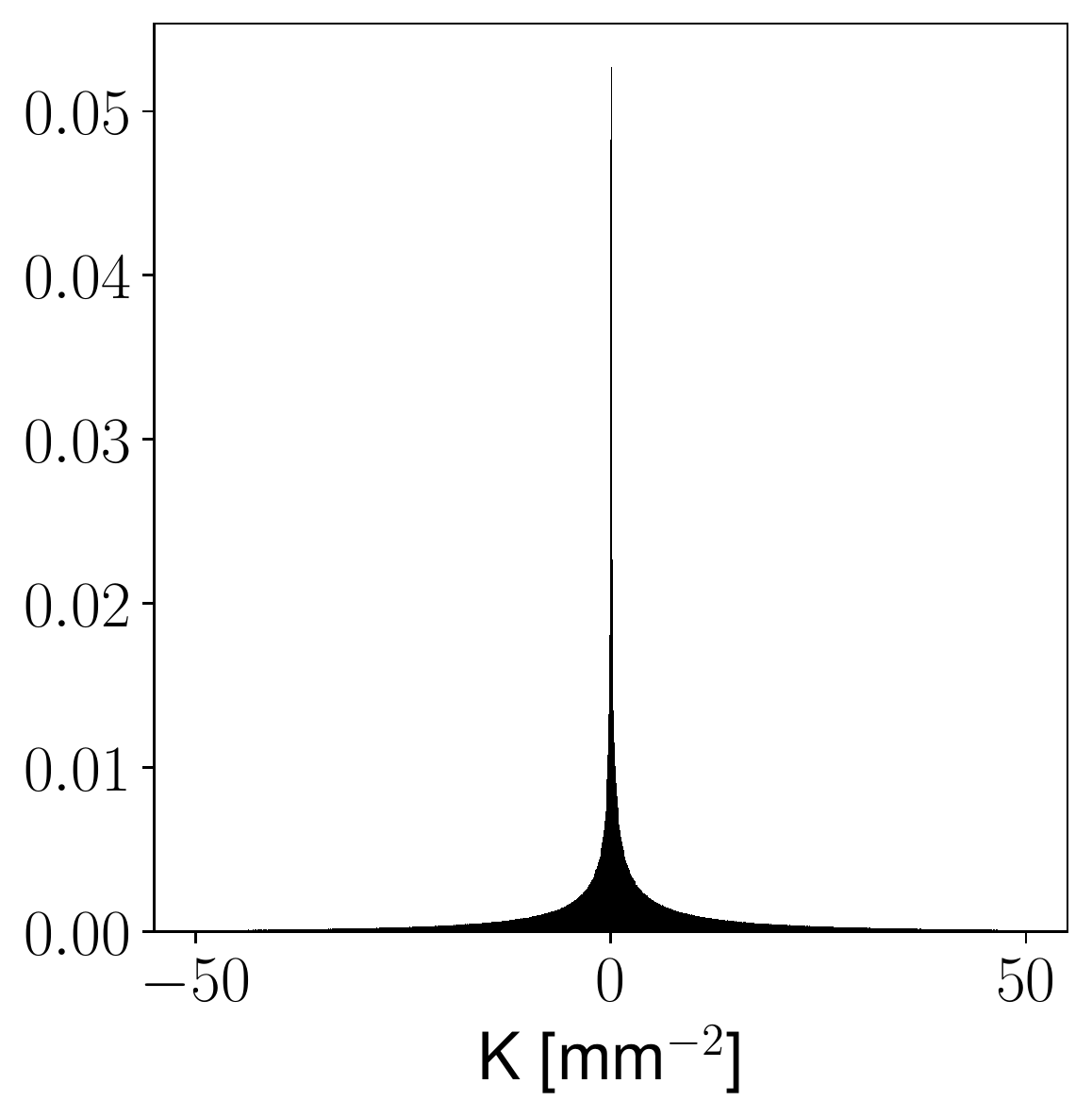}%
      \label{fig:curvature:k_hist}
    }
  \end{tabular}
  \caption{Visualizing (\ref{fig:curvature:h_surf}) mean and (\ref{fig:curvature:k_surf}) Gaussian curvature across the bone surface. Histograms of (\ref{fig:curvature:h_surf}) mean and (\ref{fig:curvature:k_surf}) Gaussian curvature show no signs of quantization.}
  \label{fig:curvature}
\end{figure*}

The filtered image and constructed image are shown in Figure~\ref{fig:density}.
Noise has been removed due to replacing the phases with their average values.
Similarly, because a regularized Heaviside function is used and the high-order signed distance transform has gradients, the constructed image has gradients at the edge between the phases.
However, the constructed image has a thicker appearance than the original image with gradients not matching local hypo-intensities in the bone phase.
This suggests the regularization parameter $\epsilon$ could be included in optimization.
It should be noted that the zero-crossings of $\psi$, $\phi$, and the constructed image correspond (Figure~\ref{fig:outline}), so the object is not actually thicker.
The object just appears thicker because of discrepancies in threshold, local intensities, and the regularization parameter.

A regression and Bland-Altman plot between the volumetric bone minearal density of the original ($\psi$) and constructed image are given in Figure~\ref{fig:agreement}.
The regression slope was $1.004~[1.002, 1.006]$ and regression intercept was $0.912~[0.480, 1.344]$, reported mean [95\% confidence intervals].
The Bland-Altman bias was 1.627 with 95\% limits of agreement of $[1.154, 2.100]$.
A small proportional bias is seen in the Bland-Altman analysis ($R^2 = 66.1\%$) caused by selecting the same density threshold independent of the object densities.
Both the bias and proportional bias is tiny $(1.41\% - 0.69\%)$ in comparison to the volumetric bone mineral density.
The component estimation method of Equations~\ref{eqn:rho1} and \ref{eqn:rho2} has the property that the constructed volume will always agree with the original image on average density.
As a result, vBMD errors originate in the ground truth data not matching the assumptions of a biphasic material, either because of noise or that the phases do not exhibit a single density.


\subsubsection{Morphometry}
\label{subsubsec:morphometry}
Regression and Bland-Altman plots are displayed in Figure~\ref{fig:morph}.
Agreement in bone volume fraction (BV/TV) is very high suggesting very few voxels were removed by component labelling.
A slight proportional bias is seen, consistent with the proportional bias in vBMD (Figure~\ref{fig:density}).
Agreement in bone surface area (BS) is high ($R^2 = 100.0\%$) with a small bias of $2.03~[-29.35, 33.42]$ (95\% LoA).
The proposed method underestimated surface area, consistent with a sub-voxel localization of the surface that is smaller than the binarized volume.
Excellent agreement ($R^2 = 100.0\%$) and small bias ($0.02~[-0.06, 0.10]$) is seen in structure model index (SMI).
The standard method estimates SMI using a finite sized dilation of the surface on the order of the sample spacing.
This appears to result in an overestimation of SMI relative to the infinitesimal definition of the proposed method.
Finally, Euler-Poincar\'e characteristic ($\chi$) shows reduced agreement ($R^2 = 96.3\%$) with a modest bias  of $-1857.78~[-3241.42, -474.1265]$.
The standard method requires component labelling to estimated $\chi$, removing many small components which would add to the total Euler-Poincar\'e characteristic.
As a result, the proposed method overestimated $\chi$ relative to the standard method.

Mean and Gaussian curvature are displayed across the surface of a single sample in Figure~\ref{fig:curvature}.
Rendered mean curvature is smooth across the surface, showing negative and positive curvature across rods and plates.
The mean curvature histogram shows a spike around zero where the image border clips the bone flat.
Rendered Gaussian curvature is less smooth, owing to the distribution being fat tailed across the surface.
The fat tailed distribution is confirmed in the histogram.
The smoothness of both the mean and Gaussian curvature agrees with the smoothness of $\psi$ in Figure~\ref{fig:surface:psi}.

\begin{figure*}
  \centering
  \begin{tabular}{ccc}
    \subfloat[Overview]{
      \includegraphics[width=0.3\linewidth]{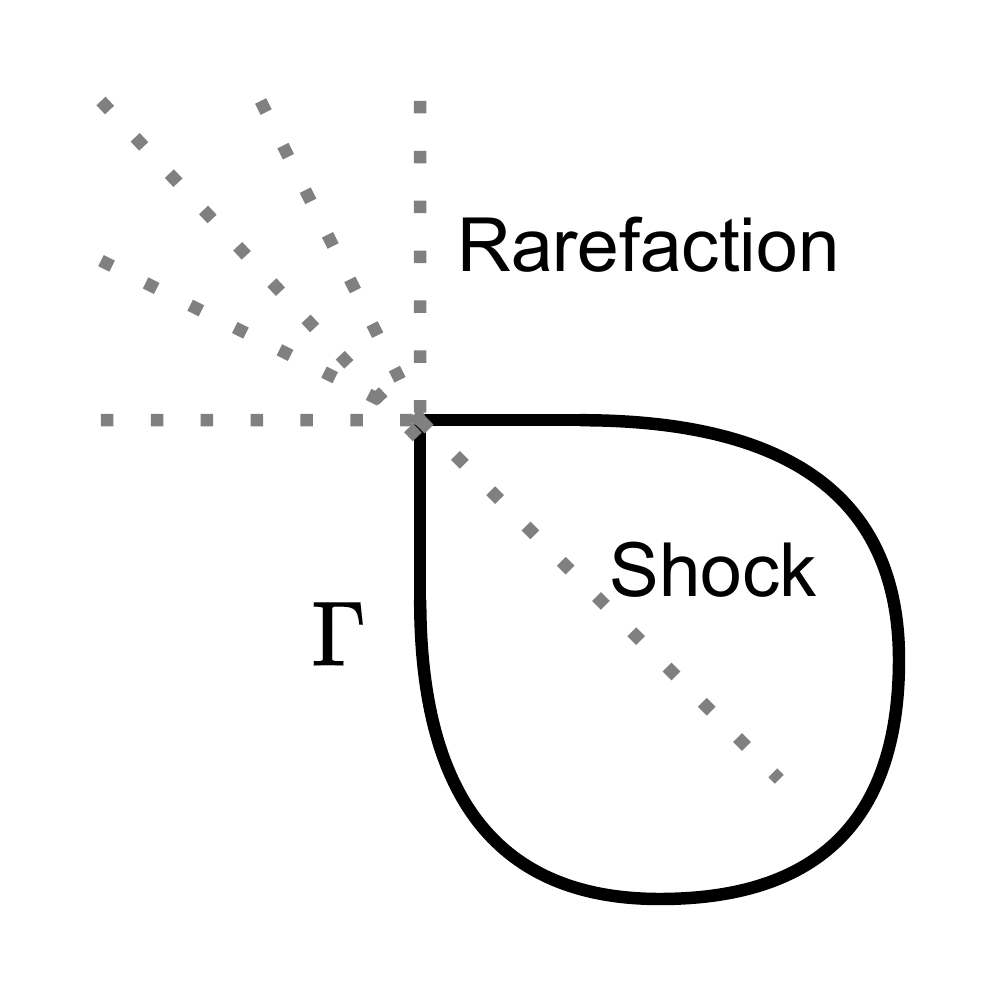}%
      \label{fig:problem:overview}
    } &
    \subfloat[Rarefaction]{
      \includegraphics[width=0.3\linewidth]{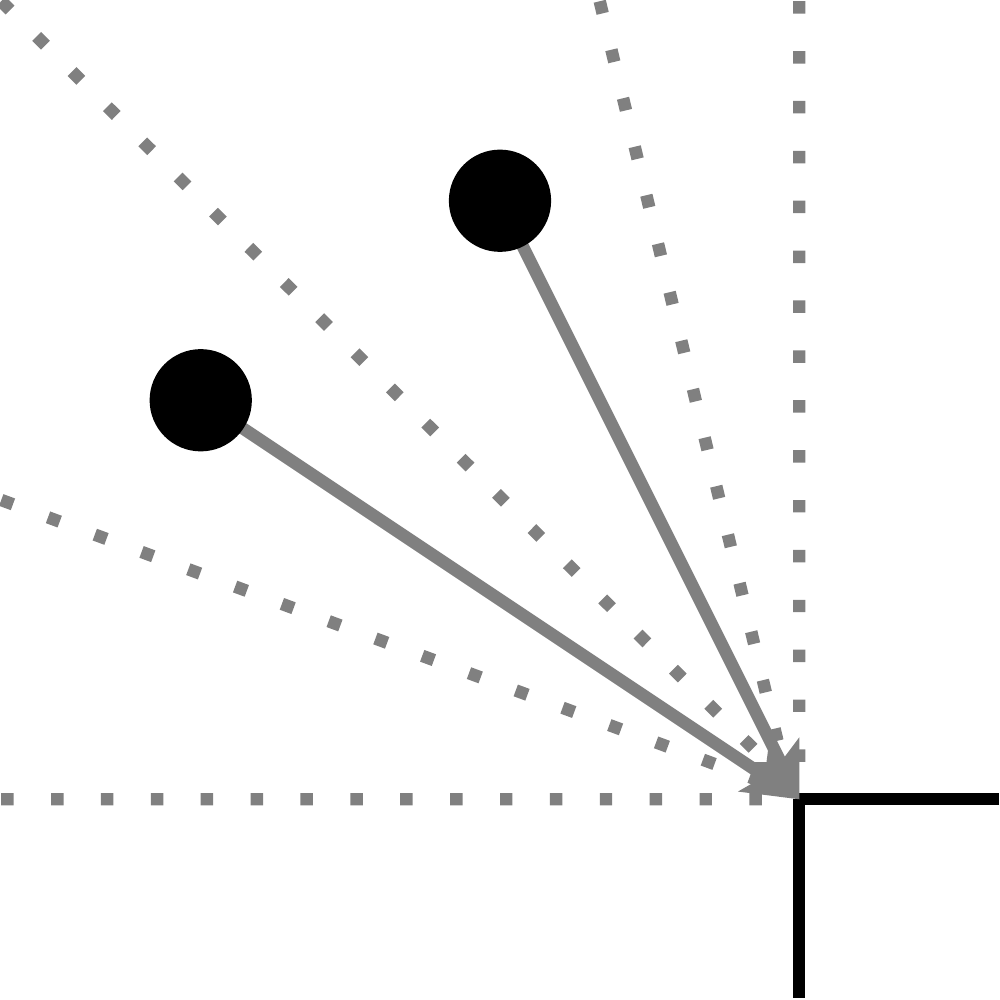}%
      \label{fig:problem:rarefaction}
    } &
    \subfloat[Shock]{
      \includegraphics[width=0.3\linewidth]{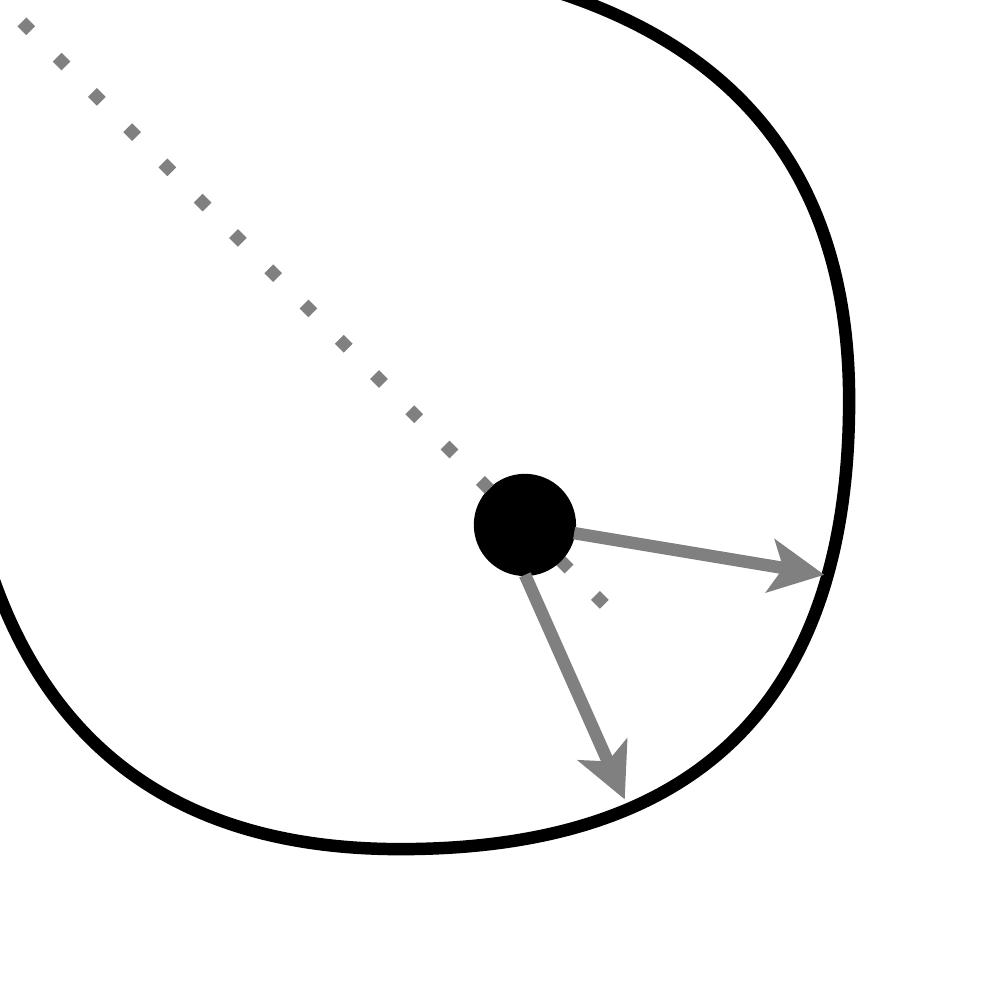}%
      \label{fig:problem:shock}
    }
  \end{tabular}
  \caption{Finding the closest point to an implicit curve. (\ref{fig:problem:rarefaction}) In areas of rarefaction, two points in the domain have a common closest point on the curve. (\ref{fig:problem:shock}) At shocks, a point in the domain has two or more closest points, consistent with Blum's grassfire burn~\cite{blum1967transformation}. All other points have a one-to-one and onto mapping to the implicit curve for a given level set. Rarefaction and shocks should be understood as regions in the related problem of simulating hyperbolic conservative laws.}
  \label{fig:problem}
\end{figure*}

\section{Discussion}
\label{sec:discussion}
A method is presented for constructing high-order signed distance transforms of two phase materials imaged with computed tomography.
The narrowband is solved by integrating voxel positions through a stationary velocity field.
The narrowband is extended to the remainder of the domain using a fast sweeping method.
The phase densities are estimated from the original density image and computed signed distance field.
The utility of the method is demonstrated by measuring bone morphometrics directly from the embedding.

The novelty of the method is to solve the narrowband from an embedding that is not the signed distance transform.
This is done by integrating narrowband voxels through a stationary velocity field until they are both sufficiently close to the zero level set and sufficiently collinear with the normal.
The modification of the convergence criterion for an image without a gradient magnitude of one was sufficient to obtain the prescribed order of accuracy.
The computation of a distance map from a grayscale image was inspiried by, but varies considerably from, the sub-pixel distance map method of Kimmel~\text{et al.}~\cite{kimmel1996sub} where a redistancing algorithm is used to construct the high-order signed distance map (called sub-pixel in that work).
Originally, the narrowband method was inspired by closest point methods~\cite{ruuth2008simple,coquerelle2016fourth} but extended for arbitrary dimensions and embeddings that are not already a signed distance field.
The same nomenclature for the $CP_\odot$ and $CP_\perp$ algorithms are used.
The narrowband algorithm is most closely related to the redistancing algorithm of Chopp~\cite{chopp2001some} with a different interpolator and modified convergence criterion.

The foundational assumption of the method was the construction of an embedding function that had the proper zero crossing.
A smoothing filter and threshold standard to bone analysis was modified to construct the implicit surface, relying on density calibration.
In general, any implicit representation will suffice as long as the zero crossing corresponds to the surface, the embedding is monotone in the narrowband, and the embedding has well-defined gradients.
However, many image analysis tasks cannot make this assumption.
Integration with alternative definitions of edges~\cite{muller1994non,prevrhal1999accuracy,treece2010high,besler2021bone} would increase the utility of the technique.
Such methods can even be directly integrated into active contour segmentation algorithms which have a penalty term forcing the embedding to be a signed distance field~\cite{li2005level}.

The limitation of the method is that the surface must be smooth, coming from smoothness assumptions in the interpolator and velocity field.
This is a good assumption for the exact task of embedding biphasic materials imaged with computed tomography where filtering has smoothed the surface.
However, the accuracy degraded at continuous but non-smooth corners of the surface.
As such, we do not recommend our method for redistancing~\cite{sussman1994level,peng1999pde,chopp2001some}.

Establishing the embedding for non-smooth surfaces is very challenging.
An attempt to alleviate the issue could be to force the divergence of the vector field to zero.
This is the case in cortical thickness mapping where Laplace's equation provides a one-to-one mapping between two surfaces~\cite{jones2000three}.
As a result, the integration paths from one surface to another cannot cross.
However, the current problem does not permit a one-to-one mapping between level sets of the embedding because there can be multiple points on the surface that are closest (in the shock regions of advection) or multiple points on the same level set that map to a single point on the surface (Figure~\ref{fig:problem}).

An improvement to the method would be to estimate the regularization parameter, $\epsilon$, from the image data.
One approach would be to estimate $\epsilon$ in conjunction with $\rho_1$ and $\rho_2$ based on a non-linear least squares.
It is expected that the magnitude of the regularization parameter corresponds to the size of the point spread function of the imager.
Such a result would improve the accuracy of the gradients in the constructed image.
Additionally, the method could be fully automated by including a threshold estimation step.
Otsu's method~\cite{otsu1979threshold} is particularly appropriate for biphasic materials, and would give a distribution instead of point estimate for phase densities.

\section{Conclusion}
\label{sec:conclusion}
A method is presented for computing high-order signed distance transforms of biphasic materials imaged with computed tomography.
Phase densities and general morphometry can be estimated from these embeddings.
The algorithm is high-order, free of the quantization artifact associated with distance transforms of sampled binary images.


%


\appendices

\appendix[Approximating Distance from the Surface]
\label{app:approx}
It is desirable to be able to select the convergence criterion based on a distance from the surface.
This allows measuring the convergence of $CP_\odot$ in physical units, permitting one to prescribe an order of accuracy.

\begin{figure}
  \centering
  \includegraphics[width=0.9\linewidth]{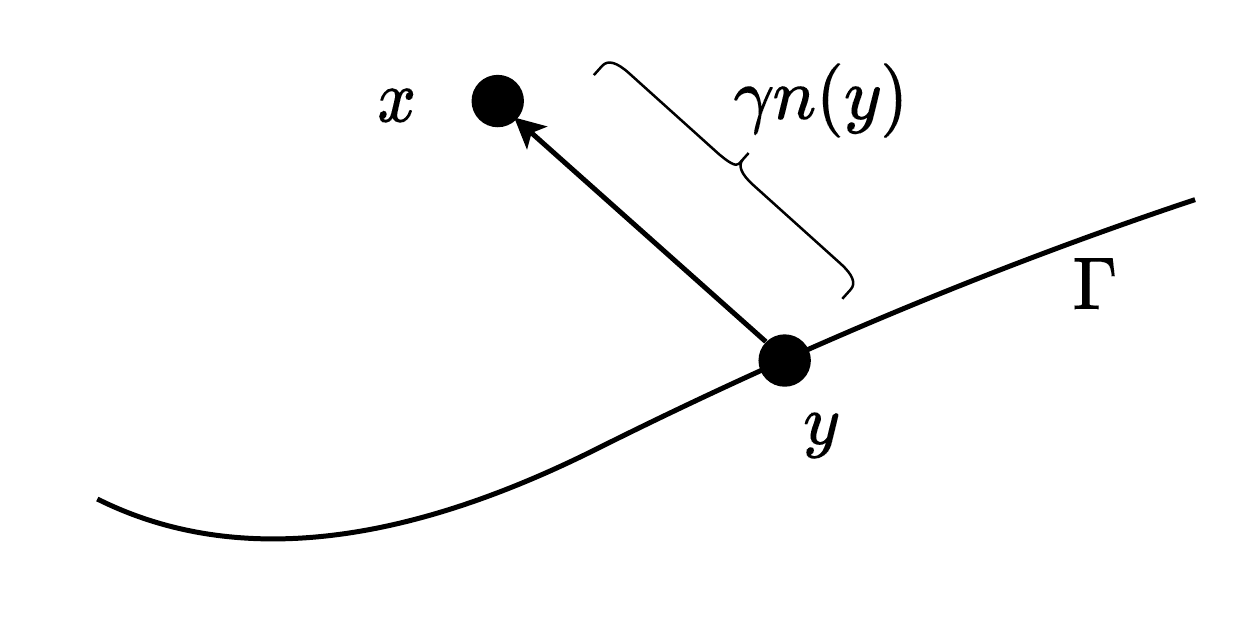}%
  \caption{Setup for estimating the distance from the embedding.}
  \label{fig:distance}
\end{figure}

The geometric picture is given in Figure~\ref{fig:distance}.
A Taylor series approximation to the implicit contour is made at the zero crossing:
\begin{equation}
  \psi(x + \delta x) = \psi(x) + \delta x^T \nabla \psi(x) + \mathcal{O}(\lVert \delta x \rVert^2)
\end{equation}
where $\delta x$ is a directional vector.
The direction is selected as the normal vector of a prescribed length:
\begin{equation}
  \delta x = \gamma n(x)
\end{equation} 
where $\gamma$ is the length.
Second, we know $\psi(x) = 0$ by definition of the zero crossing.
As such, we get the following approximation for the distance from the contour:
\begin{eqnarray}
  \psi(x + \delta x) & \approx & \gamma \frac{\nabla \psi}{|\nabla \psi|} \nabla \psi(x) \\
  \gamma & \approx & \frac{\psi(x + \delta x)}{|\nabla \psi|}
\end{eqnarray}
We modify the convergence criterion to be:
\begin{equation}
  \label{eqn:cp_dot_convergence}
  |\psi| < \gamma |\nabla \psi|
\end{equation}
This exact expression avoids division by zero in flat areas of $\psi$.
Furthermore, it prevents early stopping away from the interface since $|\psi|$ is greater than zero but $|\nabla \psi|$ approaches zero near the medial axis.
Additionally, the expression $|\nabla \psi|$ is already available from the computation of the normal in Algorithm~\ref{algo:cp}, avoiding any additional computation.
Finally, the error in the approximation decreases as the point approaches the surface at a rate $\mathcal{O}(\gamma^2)$.

If $\psi$ is a signed distance function, Equation~\ref{eqn:cp_dot_convergence} becomes the same convergence criterion as before since the gradient magnitude is unity.
In general, $|\nabla \psi|$ can be seen as a normalizing factor bringing the gradient magnitude close to one such that the distance travelled in the embedding is the same as the distance travelled in the domain.



\bibliographystyle{IEEEtran}
\bibliography{IEEEabrv,EmbedCT.bib}

\end{document}